\documentclass{article}

\usepackage{arxiv}
\usepackage{amsmath}
\usepackage{amsthm}
\usepackage{booktabs}
\usepackage{latexsym} 
\usepackage{balance}
\usepackage{dsfont}
\usepackage[geometry]{ifsym}
\usepackage{algorithm}
\usepackage{algpseudocode}
\algrenewcommand\algorithmicrequire{\textbf{Input:}}
\algrenewcommand\algorithmicensure{\textbf{Output:}}

\usepackage{multirow}
\usepackage{siunitx} 
\usepackage{tikz}
\usetikzlibrary{trees, positioning, shapes.geometric, calc, backgrounds}
\usepackage{pgfplots}
\usepackage{pgf}

\usepackage{listings}
\definecolor{palecyan}{RGB}{204, 238, 255}
\definecolor{palegreen}{RGB}{204, 221, 170}
\definecolor{palered}{RGB}{255, 204, 204}

\usepackage{hyperref}
\usepackage[utf8]{inputenc}
\usepackage[T1]{fontenc}
\usepackage{amsfonts}

\theoremstyle{definition}
\newtheorem{definition}{Definition}[section]
\newtheorem{conjecture}{Conjecture}[section]

\theoremstyle{plain}

\theoremstyle{remark}

\usepackage{subcaption} 
\usepackage{enumitem}
\usepackage{relsize}

\makeatletter
\def\pmb@@#1#2#3{\leavevmode\setboxz@h{#3}%
\dimen@-\wdz@
\kern-.5\ex@\copy\z@
\kern\dimen@\kern.25\ex@\raise.4\ex@\copy\z@
\kern\dimen@\kern.2\ex@\raise.3\ex@\copy\z@
\kern\dimen@\kern.15\ex@\raise.2\ex@\copy\z@
\kern\dimen@\kern.25\ex@\box\z@
}
\makeatother

\def\fullouterjoin{\raisebox{0.2ex}{\hspace{0.1em}\resizebox{1.2em}{1ex}{\textbf{{\tiny \pmb{\textifsym{d|><|d}}}}}}\hspace{0.1em}}

\def\ojoin{\setbox0=\hbox{$\Join$}%
\rule[0.08ex]{.28em}{.4pt}\llap{\rule[1.022ex]{.28em}{.4pt}}}
\def\fullouterjoin{\mathbin{\ojoin\mkern-6.8mu\Join\mkern-6.8mu\ojoin}}

\AtBeginDocument{%
  \providecommand\BibTeX{{%
    \normalfont B\kern-0.5em{\scshape i\kern-0.25em b}\kern-0.8em\TeX}}}

\makeatletter
\date{June 2025}           \let\Date\@date
\makeatother

\title{Sketched Sum-Product Networks for Joins}

\newif\ifuniqueAffiliation
\uniqueAffiliationtrue

\ifuniqueAffiliation 
\author{Brian Tsan,  Abylay Amanbayev, Asoke Datta, Florin Rusu\\
	University of California Merced\\
	\texttt{\{btsan, amanbayev, adatta2, frusu\}@ucmerced.edu} \\
}
\hypersetup{
pdftitle={Sketched Sum-Product Networks for Joins},
pdfauthor={Brian Tsan,  Abylay Amanbayev, Asoke Datta, Florin Rusu},
pdfkeywords={Sketches, Sum-Product Networks, Selection predicates, Multi-way join cardinality estimation}
}

\begin{document}
\maketitle

\begin{abstract}
Sketches have shown high accuracy in multi-way join cardinality estimation, a critical problem in cost-based query optimization.
Accurately estimating the cardinality of a join operation --- analogous to its computational cost --- allows the optimization of query execution costs in relational database systems.
However, although sketches have shown high efficacy in query optimization, they are typically constructed specifically for predefined selections in queries that are assumed to be given a priori, hindering their applicability to new queries.
As a more general solution, we propose for Sum-Product Networks to dynamically approximate sketches on-the-fly.
Sum-Product Networks can decompose and model multivariate distributions, such as relations, as linear combinations of multiple univariate distributions.
By representing these univariate distributions as sketches, Sum-Product Networks can combine them element-wise to efficiently approximate the sketch of any query selection.
These approximate sketches can then be applied to join cardinality estimation.
In particular, we implement the Fast-AGMS and Bound Sketch methods, which have successfully been used in prior work, despite their costly construction.
By accurately approximating them instead, our work provides a practical alternative to apply these sketches to query optimization. 
\end{abstract}

\keywords{Sketches, Sum-Product Networks, Selection predicates, Multi-way join cardinality estimation}

\section{Introduction}

Join cardinality estimation is a challenging problem in database query optimization~\cite{JOB, stats-ceb}, especially when queries include filter conditions on the joining tables~\cite{vengerov}. Traditional data synopses such as histograms and samples~\cite{synopses} are built over entire relations before querying. While this maximizes the synopses' generality, integrating filters at query time considerably degrades accuracy~\cite{vengerov}. An alternative is to evaluate the filters as a preprocessing step during query optimization and build the synopses only over the tuples satisfying the filter conditions~\cite{pessimistic}. Single-pass streaming synopses such as (hash-based) sketches~\cite{count-min,count-sketch,fast-agms} suit this approach because of their constant update time. To achieve reasonable overhead during query optimization, extensive parallelization, and even GPU acceleration may be employed~\cite {compass}.

\makeatletter
\tikzset{
    database/.style={
        path picture={
            \fill[white] (0, 1.5*\database@segmentheight) circle [x radius=\database@radius,y radius=\database@aspectratio*\database@radius];
            \fill[white] (-\database@radius, 0.5*\database@segmentheight) arc [start angle=180,end angle=360,x radius=\database@radius, y radius=\database@aspectratio*\database@radius];
            \fill[white] (-\database@radius,-0.5*\database@segmentheight) arc [start angle=180,end angle=360,x radius=\database@radius, y radius=\database@aspectratio*\database@radius];
            \fill[white] (-\database@radius,1.5*\database@segmentheight) -- ++(0,-3*\database@segmentheight) arc [start angle=180,end angle=360,x radius=\database@radius, y radius=\database@aspectratio*\database@radius] -- ++(0,3*\database@segmentheight);
            \draw (0, 1.5*\database@segmentheight) circle [x radius=\database@radius,y radius=\database@aspectratio*\database@radius];
            \draw (-\database@radius, 0.5*\database@segmentheight) arc [start angle=180,end angle=360,x radius=\database@radius, y radius=\database@aspectratio*\database@radius];
            \draw (-\database@radius,-0.5*\database@segmentheight) arc [start angle=180,end angle=360,x radius=\database@radius, y radius=\database@aspectratio*\database@radius];
            \draw (-\database@radius,1.5*\database@segmentheight) -- ++(0,-3*\database@segmentheight) arc [start angle=180,end angle=360,x radius=\database@radius, y radius=\database@aspectratio*\database@radius] -- ++(0,3*\database@segmentheight);
        },
        minimum width=2*\database@radius + \pgflinewidth,
        minimum height=3*\database@segmentheight + 2*\database@aspectratio*\database@radius + \pgflinewidth,
    },
    database segment height/.store in=\database@segmentheight,
    database radius/.store in=\database@radius,
    database aspect ratio/.store in=\database@aspectratio,
    database segment height=0.1cm,
    database radius=0.25cm,
    database aspect ratio=0.35,
}
\makeatother

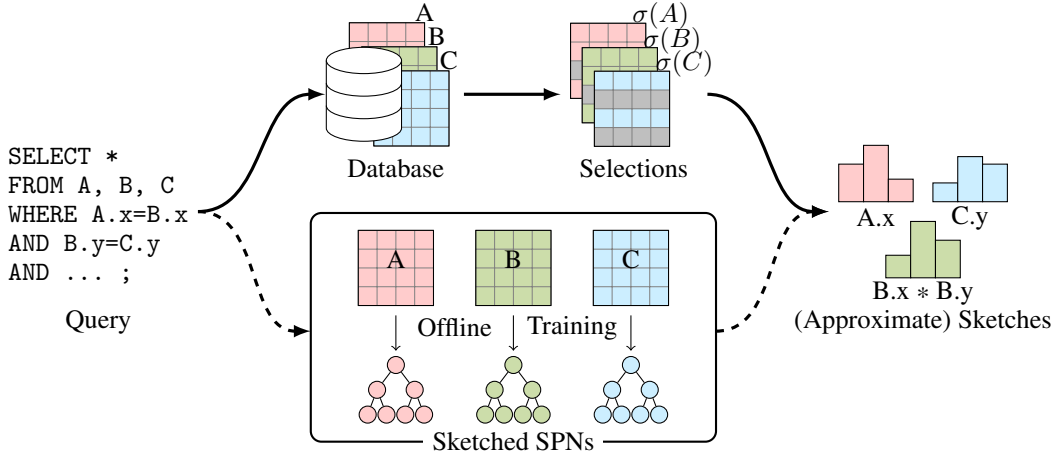
\begin{figure*}[ht]
    \centering
    \begin{tikzpicture}
        [level distance = 0.33cm,
        level 1/.style = {sibling distance=0.5cm},
        level 2/.style = {sibling distance=0.25cm},
        level 3/.style = {sibling distance=0.25cm},
        edge from parent/.style = {draw, -}]
        \node (sql) {
            \begin{lstlisting}
SELECT *
FROM A, B, C
WHERE A.x=B.x
AND B.y=C.y
AND ... ;
            \end{lstlisting}
        };
        
        \node[anchor=west] at ($(sql.east) + (2, -0.75)$) (table_A){
            \begin{tikzpicture}
                \fill[palered] (0,0) rectangle (1,1);
                \draw[step=.25cm,gray,very thin] (0,0) grid (1,1);
                \draw (0,0) rectangle (1,1);
                \node[anchor=base] at (0.5,0.5) {A};
            \end{tikzpicture}
        };
        \node[anchor=west] at ($(table_A.east) + (0.3, 0)$) (table_B){
            \begin{tikzpicture}
                \fill[palegreen] (0,0) rectangle (1,1);
                \draw[step=.25cm,gray,very thin] (0,0) grid (1,1);
                \draw (0,0) rectangle (1,1);
                \node[anchor=base] at (0.5,0.5) {B};
            \end{tikzpicture}
        };
        \node[anchor=west] at ($(table_B.east) + (0.3, 0)$) (table_C){
            \begin{tikzpicture}
                \fill[palecyan] (0,0) rectangle (1,1);
                \draw[step=.25cm,gray,very thin] (0,0) grid (1,1);
                \draw (0,0) rectangle (1,1);
                \node[anchor=base] at (0.5,0.5) {C};
            \end{tikzpicture}
        };

        \node[draw, circle, fill=palered, inner sep=0.08cm] at ($(table_A.south) + (0, -0.66)$) (spn_A){}
            child {node[draw, circle, fill=palered, inner sep=0.08cm] {}
                child {node[draw, circle, fill=palered, inner sep=0.08cm] {}}
                child {node[draw, circle, fill=palered, inner sep=0.08cm] {}}
            }
            child {node[draw, circle, fill=palered, inner sep=0.08cm] {}
                child {node[draw, circle, fill=palered, inner sep=0.08cm] {}}
                child {node[draw, circle, fill=palered, inner sep=0.08cm] {}}
            }
        ;
        \node[draw, circle, fill=palegreen, inner sep=0.08cm] at ($(table_B.south) + (0, -0.66)$) (spn_B){}
            child {node[draw, circle, fill=palegreen, inner sep=0.08cm] {}
                child {node[draw, circle, fill=palegreen, inner sep=0.08cm] {}}
                child {node[draw, circle, fill=palegreen, inner sep=0.08cm] {}}
            }
            child {node[draw, circle, fill=palegreen, inner sep=0.08cm] {}
                child {node[draw, circle, fill=palegreen, inner sep=0.08cm] {}}
                child {node[draw, circle, fill=palegreen, inner sep=0.08cm] {}}
            }
        ;
        \node[draw, circle, fill=palecyan, inner sep=0.08cm] at ($(table_C.south) + (0, -0.66)$) (spn_C){}
            child {node[draw, circle, fill=palecyan, inner sep=0.08cm] {}
                child {node[draw, circle, fill=palecyan, inner sep=0.08cm] {}}
                child {node[draw, circle, fill=palecyan, inner sep=0.08cm] {}}
            }
            child {node[draw, circle, fill=palecyan, inner sep=0.08cm] {}
                child {node[draw, circle, fill=palecyan, inner sep=0.08cm] {}}
                child {node[draw, circle, fill=palecyan, inner sep=0.08cm] {}}
            }
        ;
        \draw[-{>[sep=2pt]}] (table_A.south) -- (spn_A.north);
        \draw[-{>[sep=2pt]}] (table_B.south) -- (spn_B.north);
        \draw[-{>[sep=2pt]}] (table_C.south) -- (spn_C.north);

        \node at ($(table_A.south)!0.5!(table_B.south) + (0, -0.16)$) {Offline};
        \node at ($(table_B.south)!0.5!(table_C.south) + (0, -0.16)$) {Training};

        \node[anchor=south] at ($(table_A.north) + (0, 1.5)$) (table_A1){
            \begin{tikzpicture}
                \fill[palered] (0,0) rectangle (1,1);
                \draw[step=.25cm,gray,very thin] (0,0) grid (1,1);
                \draw (0,0) rectangle (1,1);
                \node at (1,0.9) {A};
            \end{tikzpicture}
        };
        \node[anchor=south] at ($(table_A.north) + (0.16, 1.5-0.32)$) (table_B1){
            \begin{tikzpicture}
                \fill[palegreen] (0,0) rectangle (1,1);
                \draw[step=.25cm,gray,very thin] (0,0) grid (1,1);
                \draw (0,0) rectangle (1,1);
                \node at (1,0.9) {B};
            \end{tikzpicture}
        };
        \node[anchor=south] at ($(table_A.north) + (0.32 , 1.5-0.64)$) (table_C1){
            \begin{tikzpicture}
                \fill[palecyan] (0,0) rectangle (1,1);
                \draw[step=.25cm,gray,very thin] (0,0) grid (1,1);
                \draw (0,0) rectangle (1,1);
                \node at (1,0.9) {C};
            \end{tikzpicture}
        };
        \node at (table_C1.west) (db1) [database, database radius=0.5cm, database segment height=0.3cm] {};

        \node at ($(table_C.north |- table_A1.east) + (0.04, 0)$) (sigma_A){
            \begin{tikzpicture}
                \fill[palered] (0,0) rectangle (1,1);
                \fill[lightgray] (0,0.25) rectangle (1,0.5);
                \draw[step=.25cm,gray,very thin] (0,0) grid (1,1);
                \draw (0,0) rectangle (1,1);
                \node at (1.2,1.1) {$\sigma(A)$};
            \end{tikzpicture}
        };
        \node at ($(table_C.north |- table_B1.east) + (0.2, 0)$) (sigma_B){
            \begin{tikzpicture}
                \fill[palegreen] (0,0) rectangle (1,1);
                \fill[lightgray] (0,0.25) rectangle (1,0.5);
                \draw[step=.25cm,gray,very thin] (0,0) grid (1,1);
                \draw (0,0) rectangle (1,1);
                \node at (1.2,1.1) {$\sigma(B)$};
            \end{tikzpicture}
        };
        \node at ($(table_C.north |- table_C1.east) + (0.36, 0)$) (sigma_C){
            \begin{tikzpicture}
                \fill[palecyan] (0,0) rectangle (1,1);
                \fill[lightgray] (0,0.5) rectangle (1,0.75);
                \fill[lightgray] (0,0) rectangle (1,0.25);
                \draw[step=.25cm,gray,very thin] (0,0) grid (1,1);
                \draw (0,0) rectangle (1,1);
                \node at (1.2,1.1) {$\sigma(C)$};
            \end{tikzpicture}
        };

        \node[anchor=south west] at ($(table_C.east |- sql.east) + (2, 0)$) (sketch_A) {
            \begin{tikzpicture}
                \draw[fill=palered] (0,0) rectangle (0.33,0.5);
                \draw[fill=palered] (0.33,0) rectangle (0.66,0.75);
                \draw[fill=palered] (0.66,0) rectangle (0.99,0.3);
            \end{tikzpicture}
        };
        \node at ($(sketch_A.south) + (0, -0.08)$) {A.x};
        \node [anchor=north] at ($(sketch_A.east |- sketch_A.south)$) (sketch_B) {
            \begin{tikzpicture}
                \draw[fill=palegreen] (0,0) rectangle (0.33,0.3);
                \draw[fill=palegreen] (0.33,0) rectangle (0.66,0.75);
                \draw[fill=palegreen] (0.66,0) rectangle (0.99,0.5);
            \end{tikzpicture}
        };
        \node at ($(sketch_B.south) + (0, -0.08)$) {B.x \textasteriskcentered{} B.y};
        \node [anchor=south west] at ($(sketch_B.north |- sketch_A.south)$) (sketch_C) {
            \begin{tikzpicture}
                \draw[fill=palecyan] (0,0) rectangle (0.33,0.25);
                \draw[fill=palecyan] (0.33,0) rectangle (0.66,0.6);
                \draw[fill=palecyan] (0.66,0) rectangle (0.99,0.5);
            \end{tikzpicture}
        };
        \node at ($(sketch_C.south) + (0, -0.08)$) {C.y};

        \node at ($(sql.south) + (0, -0.32)$) (step1) {Query};
        \node at ($(sketch_B.south |- step1.east)$) (step3) {(Approximate) Sketches};

        \node at ($(table_A1.south)  + (0, -0.64-0.16)$) (step2_old) {Database};
        \node at ($(sigma_A.south |- step2_old.east)$) (step3_old) {Selections};

        \draw[-{latex}, very thick] (sql.east) to [out=0, in=180] (db1.west);
        \draw[-{latex}, very thick] ($(table_C1.east |- db1.east) - (0.16, 0)$) -- ($(sigma_A.west |- db1.east)$);
        \draw[-{latex}, very thick] ($(sigma_C.east |- db1.east) - (0.32, 0)$) to [out=0, in=180] ($(sketch_A.west |- sketch_B.north)$);
        
        \draw[-{latex}, very thick, dashed] (sql.east) to [out=0, in=180] ($(table_A.west |- spn_A.north) + (-0.5, 0.33)$);
        \draw[-{latex}, very thick, dashed] ($(table_C.east |- spn_A.north) + (0.5, 0.33)$) to [out=0, in=180] ($(sketch_A.west |- sketch_B.north)$);

        \begin{pgfonlayer}{background}
            \draw[rounded corners, thick] ($(table_A.north west) + (-0.5, 0.1)$) rectangle ($(table_C.east |- spn_C.south) + (0.5, -0.9)$);
        \end{pgfonlayer}
        \node at ($(spn_B.south) + (0, -0.9)$) [fill=white]{Sketched SPNs};
    \end{tikzpicture}
    \caption{Comparison of sketching pipelines.
    Given a query, previous methods must first filter the relations to compute the selections to sketch.
    The proposed approximate sketching pipeline (dashed arrows) uses Sum-Product Networks to approximate sketches without computing selections.
    Training can be completed offline.}
    \label{fig:overview}
\end{figure*}

In recent work~\cite{ApproximateSketches}, bidirectional transformers~\cite{bert} have been used to infer the sketches of query selections. This approximates --- rather than builds from scratch --- the necessary sketches to estimate the cardinality of joins with filter conditions. Thus, it avoids the construction overhead altogether. However, relying on deep learning constrains the scalability of the approximate sketches. The number of trainable parameters increases linearly with the size of the sketch, thus only sketches with relatively small width can be trained, due to the limited memory capacity of hardware accelerators, e.g., GPU.
This is a significant shortcoming since the width of the sketch correlates with its accuracy.

To overcome this lack of scalability to larger sketches, we propose Sketched Sum-Product Networks --- or Sketched SPNs.
Unlike deep learning models, Sum-Product Networks (SPNs) are not reliant on hardware acceleration for training.
Hence, they can be used to approximate larger sketches, which results in more accurate cardinality estimation and effective query optimization.
Furthermore, SPNs have been shown~\cite{mspn} to be applicable to multimodal data --- both discrete and continuous types --- which is required to model relations with multiple attribute types.

Our main idea is to store sketches in an SPN's leaf nodes --- which each represent a partition ---  and combine these sketches over the structure of the SPN.
Filter conditions are also applied to the sketches during approximation, such that the resulting sketch obtained at the root of the SPN is an approximate sketch of their selection.
We primarily consider the Fast-AGMS sketch~\cite{fast-agms}, which is an accurate cardinality estimator for multi-way joins~\cite{conv-sketch}.
Our method also generalizes to other sketches.
In particular, our open-source implementation~\cite{github} includes Bound Sketch~\cite{pessimistic}, which is a \textit{pessimistic}~\cite{pessimistic-cardest} estimator that gives upper bounds join cardinality with high efficacy for query optimization.

Our detailed technical contributions are as follows:
\begin{itemize}[leftmargin=*]
\item Sketched Sum-Product Networks, a practical method for approximating sketches as an alternative pipeline to applying filter conditions before sketching, illustrated in \autoref{fig:overview}.

\item An upper bound on the approximation error of sketches by SPNs, which we verify in our experiments.

\item The application of Bound Sketch with cross-correlation~\cite{conv-sketch} to avoid exponential space complexity.

\item An upward-biased estimate derived from Fast-AGMS that outperforms other estimators in query optimization.

\item Sketched SPNs perform within $3\%$ of the fastest query execution time for the JOB-light and Stats-CEB workloads.
\end{itemize}

\section{Problem Definition}

In relational database systems, cost-based query optimizers often estimate the cost of potential query execution plans as a function of their size or cardinality, defined as the number of tuples a query would return.
Cardinality is a close analog to the query's actual computational runtime.
We address the problem of estimating the cardinality of joins, which may include a variable number of relations, subject to selection predicates.

\subsection{Multi-way Join Cardinality Estimation}
In a two-way equi-join, the join cardinality between two relations is the number of pairs of tuples between both relations whose values for a given join attribute are equivalent.
Consider two relations, $T_1$ and $T_2$, which join on their respective join attributes with the domain $I$.
The cardinality of this join is defined as follows:
\begin{equation}
    \left| T_1 \Join T_2 \right| = \sum_{i \in I} f_1(i) f_2(i)
    \label{eq:join-size-estimate}
\end{equation}
\noindent
where $f_1(i) $ and $f_2(i)$ denote the frequencies of the join attribute element $i$ in $T_1$ and $T_2$, respectively.
The extension to multi-way joins is by introducing additional frequencies for every relation.
Let $\{T_1, \dots, T_n\}$ be the relations in an $n$-ary join on attributes that share the domain $I$, i.e., each relation has a single join attribute.
In this case, the join cardinality is the following sum of products:
\begin{equation}
    \left| T_1 \Join \dots \Join T_n \right| = \sum_{i \in I}
        f_1(i) \cdots f_n(i)
\end{equation}
\noindent
where $f_k(i)$ denotes the frequency of the join attribute element $i$ in the $k$-th relation. In practice, the exact frequencies $f_k(i)$ are unknown and may be estimated using histograms~\cite{histograms}, samples~\cite{vengerov}, sketches~\cite{Rusu:tods-2008:sketches-join-size}, or other synopses~\cite{synopses}.

In this paper, we consider the general case in which every relation may have multiple join attributes with different domains.
We borrow the formulation by Heddes et al.~\cite{conv-sketch} and let $\{I_1, \dots, I_n\}$ denote the join domains in every relation, where $I_k$ is the cross product of the individual join attribute domains in $T_k$.
The cardinality of a multi-way join is expressed as:
\begin{equation}
    \left| T_1 \Join \dots \Join T_n \right| = \sum_{i \in \{I_1 \times \dots \times I_n\}}
        f_1(i) \cdots f_n(i) 
        \prod_{\{u, v\} \in E} \mathds{1}_{i_u = i_v} 
\end{equation}
\noindent
where $i$ is a tuple from the cross product $\{I_1 \times \cdots \times I_n\}$ and $f_k(i)$ denotes the frequency of tuples in $T_k$ that have the same value(s) for the join attribute(s) shared in $i$.
Additionally, $\{u, v\}$ are attributes joining a pair of relations in a join graph with edges $E$.
The indicator function $\mathds{1}_{i_u = i_v}$ returns 1 if the attribute value $i_u$ equals $i_v$, otherwise it returns 0, such that $\prod \mathds{1}_{i_u=i_v}$ equals 1 if and only if $i$ satisfies all the equi-joins predicates.

\subsection{Join Cardinality Subject to Filters}

Estimators that represent only the probability distribution of one random variable, a single attribute, are defined as being univariate.
Univariate estimators are challenging to apply to join cardinality estimation subject to filter conditions~\cite{vengerov}---where joins are between selections, subsets specified by filters on a relation.

With univariate estimators, frequencies may be approximated by assuming independence among attributes.
Let $f_k'(i)$ denote the frequency of the join attribute element $i$ in a selection $\sigma(T_k) \subseteq T_k$.
Assuming independence between all attributes, $f'_k(i)$ can be approximated using univariate probabilities and frequencies:
\begin{equation}
    f'_k(i) \approx f_k(i) \prod_{r \in T_k} P\left( \varphi_r \right)
\end{equation}
\noindent
where $\varphi_r \in \varphi$ denotes the predicate on attribute $r \in T_k$ and has the estimated selectivity $P(\varphi_r)\in[0,1]$---the probability that an element of $u$ satisfies $\varphi_r$.
The product of all $P\left(\varphi_r\right)$ is the joint probability that all attributes of a tuple in $T_k$ satisfy $\varphi$, by the definition of (mutual) independence~\cite{independence}.

Using $f'$, the cardinality of a multi-way join between selections can be approximated as follows:
\begin{equation}
    \label{eq:multi-selection-join}
    \left| \sigma(T_1) \Join \dots \Join \sigma(T_n) \right| =
    \sum_{i \in \{I_1 \times \dots \times I_n\}}
        f'_1(i) \cdots f'_n(i) 
        \prod_{\{u, v\} \in E} \mathds{1}_{i_u = i_v}
\end{equation}
However, this approximation may be inaccurate unless attributes are independent~\cite{Tzoumas:pvldb-2011:sel-estimation-no-indep}.

\subsection{Multivariate Estimators}

Multivariate estimators, e.g., multidimensional histograms~\cite{Deshpande:sig-record-2001:multidim-hist}, do not assume independence and may thus be more accurate, but typically require space exponential to the number of attributes.
Exceptionally, machine learning models have been shown to tractably learn joint probability distributions, even for all the many attributes of a full outer join of several relations.

The cardinality of a multi-way join between selections is proportional to the joint probability of the predicates on all full outer join attributes.
A model can estimate the probability that a tuple in the full outer join satisfies all the equi-join predicates $u=v$ with the probability of satisfying the selection predicates $\varphi_r$, for every attribute $r$ as follows:
\begin{equation}
    \left| \sigma(T_1) \Join \dots \Join \sigma(T_n) \right| \approx \widehat{P}\left( \bigwedge_{\{u, v\} \in E} u = v \bigwedge_{r \in T_1 \cup \cdots \cup T_n} \varphi_r \right) \left| T_1 \fullouterjoin \cdots \fullouterjoin T_n \right|
    \label{eq:full-outer-joint-pdf}
\end{equation}
\noindent
The join cardinality is approximated by scaling the estimated joint probability by the size of the full outer join $\left|T_1 \fullouterjoin \cdots \fullouterjoin T_n\right|$.
However, it is expensive to compute the full outer join -- or even a sample of it -- between many relations.

Instead of defining a single multivariate estimator over a full outer join, we consider \textbf{the problem of defining an  multivariate estimator over the join and filter attributes of each relation, independently}. Doing so mitigates the independence assumption of univariate estimators and avoids the training cost of computing a full outer join. Moreover, such an estimator supports the inference of the frequencies in \autoref{eq:multi-selection-join} for join cardinality estimation subject to selection predicates. The challenges of defining a multivariate estimator per relation are two-fold. First, the estimator has to support filters on any subset of attributes and sub-joins involving only a subset of the join attributes---both of which are essential in query plan enumeration. Second, the relation-level estimators must be combined into an ensemble estimator for the full multi-way join and any sub-joins. Our solution is to \textbf{use a Sum-Product Network of sketches as the multivariate estimator}, where SPNs handles the filters while sketches estimate the multi-way joins.


\section{Preliminaries}\label{prelim}

This section provides background on sketches and Sum-Product Networks, which we combine for multi-way join cardinality estimation subject to filter conditions.

\subsection{Fast-AGMS Sketch}
\label{prelim:fast-agms}

We utilize the Fast-AGMS sketch~\cite{fast-agms}, which is an unbiased frequency estimator.
The basic structure of this sketch\footnote{Also referred to as the Count Sketch~\cite{count-sketch}, which is identical, but without extension to joins. Thorup and Zhang~\cite{fast-count} also independently proposed a similar extension.} is an array of $w$ counters updated by a pair of hash functions.
The first hash function $h: \mathds{R} \rightarrow \left\{1, ..., w\right\}$ maps a given element to a counter.
The other hash function $\xi: \mathds{R} \rightarrow \pm 1$ determines whether to increment or decrement that counter.
In this work, we refer to the dimensionality $w$ as the \textit{width} of the sketch. 

For example, consider the zero-initialized vector $a \in \mathds{R}^w$.
The Fast-AGMS update is the following:
\begin{equation}
    a_{h(x)} \leftarrow a_{h(x)} + \xi(x)
    \label{eq:sketch-update-single}
\end{equation}
\noindent
For each element $x$, e.g., from an attribute, the counter indicated by $h(x)$ is updated by $\xi(x)$.
The frequency of a specific $x$ can be approximately recovered as the product of $a_{h(x)}$ and $\xi(x)$:
\begin{equation}
    f(x) = \mathds{E}\left[ a_{h(x)} \xi(x) \right]
\end{equation}
\noindent
This is proven to be unbiased~\cite{agms} when $\xi: \mathds{R} \rightarrow \pm1$ is pairwise independent~\cite{fast_random_variables}, such that $\mathds{E}\left[\xi(x)\xi(y)\right] = 0$ if $x \neq y$.

The join cardinality of two relations can be estimated unbiasedly via the dot product of their sketches:
\begin{equation}
    \left| A \Join B \right| \approx a \cdot b
\end{equation}
\noindent
where $a$ and $b$ are the sketches of the join attributes in relations $A$ and $B$, respectively.
These corresponding sketches must share the same hash functions.

\subsubsection{Multi-Way Joins}
For multi-way joins, a distinct $\xi$ hash function is defined for each join.
Consider a three-way join $A \Join B \Join C$.
The join $A \Join B$ is assigned $\xi_a$, whereas $B \Join C$ is assigned $\xi_c$.
Since $B$ joins with both $A$ and $C$, its sketch is constructed using both $\xi_a$ and $\xi_c$:
\begin{equation}
    b_{h(x)} \leftarrow b_{h(x)} + \xi_a(x) \xi_c(x)
\end{equation}
\noindent
In contrast, sketches $a$ and $c$ are updated as in \autoref{eq:sketch-update-single} using only either $\xi_a$ or $\xi_c$, respectively.

When $A \Join B \Join C$ is a transitive join, i.e., each relation uses a single join attribute such that $A$ and $C$ both join the same attribute in $B$, then the sketch vector $b \in \mathds{R}^w$ is constructed by the same hash function $h$ as $a \in \mathds{R}^w$ and $c \in \mathds{R}^w$.
This enables the unbiased three-way join cardinality estimation via the element-wise product of all three sketches:
\begin{equation}
    \left| A \Join B \Join C \right| \approx \sum^w a \circ b \circ c
\end{equation}
\noindent
However, this only applies to transitive joins.

When $A \Join B \Join C$ is not transitive, i.e., $A$ and $C$ join on different attributes in $B$, then $b \in \mathds{R}^{w \times w}$ is a sketch matrix constructed using $h_a$ and $h_c$, which correspond with the sketch vectors $a \in \mathds{R}^w$ and $c \in \mathds{R}^w$, respectively.
The sketch matrix $b$ is constructed by the following update:
\begin{equation}
    b_{h_a(x),h_c(y)} \leftarrow b_{h_a(x),h_c(y)} + \xi_a(x) \xi_c(y)
\end{equation}
\noindent
where $x$ and $y$ are elements from two different attributes of $B$ that join with $A$ and $C$, respectively.
The three-way join cardinality estimation process now requires a matrix-vector product:
\begin{equation}
    \left| A \Join B \Join C \right| \approx a \cdot b \cdot c
\end{equation}
\noindent
In general, this extends to complex multi-way joins with sketch tensors.
The sketch of a relation in a join with $n-1$ relations would be an $(n-1)$-order tensor and the multi-way join cardinality estimated via tensor contraction.

\subsubsection{Cross-Correlation}
To avoid the exponentially large space requirements of tensors, Heddes et al.~\cite{conv-sketch} showed that (circular) cross-correlation can effectively approximate tensor contraction between Fast-AGMS sketches with just $\mathcal{O}(w)$ space, i.e., vectors.

\begin{definition}[Cross-Correlation]
    Two vectors $a, b \in \mathds{R}^w$ are cross-correlated by 
    $a \star b = \mathcal{F}^{-1}\left(\overline{\mathcal{F}a} \circ \mathcal{F}b \right)$ where $\star$ denotes the operator and $\mathcal{F}$ denotes a discrete Fourier transform.
\end{definition}

Returning to our three-way join example $A \Join B \Join C$, cross-correlation requires $b \in \mathds{R}^w$ to be a convolved sketch vector constructed by the following update:
\begin{equation}
    b_{H(x, y)} \leftarrow b_{H(x, y)} + \xi_a(x) \xi_c(y)
\end{equation}
\begin{equation}
    H(x, y) = \left(h_a(x) + h_c(y)\right) \bmod w
\end{equation}
\noindent
where $H(x, y): \mathds{R}^2 \rightarrow \left\{1, ..., w\right\}$ is a composite hash function of $h_a$ and $h_b$ used to map the join attributes $(x, y)$ to a counter.
It is equivalent to the circular convolution of two sketches containing just $x$ and $y$, respectively.
Then, cross-correlation can estimate the cardinality of the multi-way join:
\begin{align}
    \left| A \Join B \Join C \right| & \approx \sum^w a \star b \star c
    \nonumber \\ 
    & \approx \sum^w \mathcal{F}^{-1} \left( \mathcal{F}a \circ \overline{\mathcal{F}b} \circ \mathcal{F}c\right)
\end{align}

The estimation time complexity is $\mathcal{O}\left(n w\log w\right)$ time, where $n$ is the number of relations and $w \log w$ is the complexity of the Fast Fourier Transform~\cite{FFT}.
We adopt the use of cross-correlation in this work to allow for larger sketch dimensionality $w$, which also improves estimation accuracy.


\subsection{Bound Sketch}

As an alternative to Fast-AGMS, we also utilize the pessimistic join cardinality estimation method, Bound Sketch, proposed by Cai et al.~\cite{pessimistic}.
Whereas Fast-AGMS is unbiased, Bound Sketch estimates are upper bounds for the cardinality of joins, which are less likely to lead to catastrophically suboptimal plans~\cite{elephant}.

Like Fast-AGMS, the Bound Sketch uses a hash function $h: \mathds{R} \rightarrow \left\{1, ..., w\right\}$ to map elements to one of $w$ counters.
Unlike Fast-AGMS, each insertion increments a counter.
This produces the Count-Min sketch~\cite{count-min}, which is an upper bounds estimator.
However, Bound Sketch tightens the upper bound by utilizing the \textit{maximum degree} of elements inserted to a counter.
The maximum degree is defined as the largest frequency of any inserted value.

For the two-way join $A \Join B$, let $\mathcal{C}(A) \in \mathds{R}^w$ and $\mathcal{C}(B) \in \mathds{R}^w$ denote the Count-Min Sketch for the join attribute elements of relations $A$ and $B$, respectively.
Furthermore, let $\mathcal{D}(A) \in \mathds{R}^w$ and $\mathcal{D}(B) \in \mathds{R}^w$ be $w$-dimensional sketch vectors whose elements are the maximum degree of values mapped to the corresponding counters in $\mathcal{C}(A)$ and $\mathcal{C}(B)$, respectively.
The Bound Sketch upper bound is then given by the following:
\begin{equation}
    \lvert A \Join B \rvert \leq 
    \min \left\{\substack{\mathcal{C}(A) \cdot \mathcal{D}(B)\\ \mathcal{D}(A) \cdot \mathcal{C}(B)}\right\}
\end{equation}
\noindent
Both products are overestimates, hence the minimum is taken as the tighter bound.
Intuitively, each tuple in a relation (e.g., $\mathcal{C}(A)$) can only join up to the maximum degree of the other relation's join attribute (e.g., $\mathcal{D}(B)$).

For completeness, we also show the estimation for our three-way join example $A \Join B \Join C$:
\begin{equation}
    \lvert A \Join B \Join C \rvert \leq
    \min \left\{\substack{\mathcal{C}(A) \cdot \mathcal{D}(B) \cdot \mathcal{D}(C)\\
                        \mathcal{D}(A) \cdot \mathcal{C}(B) \cdot \mathcal{D}(C) \\
                        \mathcal{D}(A) \cdot \mathcal{D}(B) \cdot \mathcal{C}(C)}\right\}
\end{equation}
\noindent
In general, the extension to multi-way joins is the same as for Fast-AGMS --- tensor contraction.
However, cross-correlation can be used to approximate tensor contraction, which also allows us to apply Bound Sketch with larger sizes than prior work.

Cai et al.~\cite{pessimistic} noted that the Bound Sketch estimator has exceptionally high latency, inflating query optimization time.
This is due to its inability to estimate subject to filter conditions --- the Bound Sketch of a selection must be exactly computed by scanning and filtering its base relation.
This costly operation can even exceed the query execution time.
This is also the case for Fast-AGMS, where the accepted practice~\cite{compass, conv-sketch} has been to apply the filters just before computing the sketch.
Our proposed method uses Sum-Product Networks to approximate the sketch of the filtered relation, without necessitating a scan at estimation time.

\begin{figure*}[ht]
    \begin{subfigure}[b]{0.47\linewidth}
        \centering
        \begin{tikzpicture}[
              every node/.style = {draw, rectangle, minimum size=1cm, inner sep=2pt, align=center},
              leaf/.style = {draw, circle, minimum size=0.5cm, inner sep=2pt, align=center},
              weight/.style = {draw=none, minimum size=0, inner sep=0, align=center},
              level 1/.style = {sibling distance=3.5cm},
              level 2/.style = {sibling distance=2.5cm},
              level 3/.style = {sibling distance=1.5cm},
              edge from parent/.style = {draw, -latex}
            ]
            
            \node (root){
            \begin{tabular}{cc|c}
              $x_1$ & $y_1$ & $z_1$ \\
              $x_2$ & $y_2$ & $z_2$ \\
              $x_3$ & $y_3$ & $z_3$ \\
            \end{tabular}
            }
              child {node (sum1){
                \begin{tabular}{cc}
                  $x_1$ & $y_1$ \\
                  $x_2$ & $y_2$ \\
                  \midrule
                  $x_3$ & $y_3$ \\
                \end{tabular}
              }
                    child {node (prod1){
                    \begin{tabular}{c|c}
                        $x_1$ & $y_1$ \\
                        $x_2$ & $y_2$ \\
                    \end{tabular}
                }
                    child {node (x1)[leaf]{x}}
                    child {node (y1)[leaf]{y}}}
                child {node (prod2){
                    \begin{tabular}{c|c}
                        $x_3$ & $y_3$ \\
                    \end{tabular}
                }
                    child {node (x2)[leaf] {x}}
                    child {node (y2)[leaf] {y}}
              }}
              child {node (z)[leaf]{z}}
            ;
            
            \node at ($(sum1.south west)!.5!(prod1.north)$) [weight, above left] {$\frac{2}{3}$};
            \node at ($(sum1.south east)!.5!(prod2.north)$) [weight, above right] {$\frac{1}{3}$};
        \end{tikzpicture}
        \subcaption{Learning the Sum-Product Network structure starting from a table at the root.
        Each node recursively partitions the table either column-wise or row-wise.
        Terminates into leaf nodes containing the local univariate probability distribution of a single column.}
        \label{subfig:learn-spn}
    \end{subfigure}
    \hfill
    \begin{subfigure}[b]{0.47\linewidth}
        \centering
        \begin{tikzpicture}[
              every node/.style = {draw, circle, minimum size=1cm, inner sep=2pt, align=center},
              leaf/.style = {draw, circle, minimum size=0.5cm, inner sep=2pt, align=center},
              prob/.style = {draw=none, red, minimum size=0, inner sep=0, align=center},
              weight/.style = {draw=none, minimum size=0, inner sep=0, align=center},
              level 1/.style = {sibling distance=3.5cm},
              level 2/.style = {sibling distance=2.5cm},
              level 3/.style = {sibling distance=1.5cm},
              sum/.style = {draw, circle, very thick, minimum size=0.7cm, inner sep=2pt, align=center,
            path picture={\draw
                   (path picture bounding box.south) -- (path picture bounding box.north)
                   (path picture bounding box.west) -- (path picture bounding box.east);}},
               product/.style = {draw, circle, very thick, minimum size=0.7cm, inner sep=2pt, align=center, path picture={\draw
                   (path picture bounding box.south east) -- (path picture bounding box.north west)
                   (path picture bounding box.south west) -- (path picture bounding box.north east);}},
              edge from parent/.style = {draw, -}
            ]

            \node [product](root){}
              child {node (sum1)[sum]{}
                    child {node (prod1)[product]{}
                    child {node (x1)[leaf]{x}}
                    child {node (y1)[leaf]{y}}}
                child {node (prod2)[product]{}
                    child {node (x2)[leaf]{x}}
                    child {node (y2)[leaf]{y}}
              }}
              child {node (z)[leaf]{z}}
            ;

            \draw[densely dotted, -latex, red] (z.north) .. controls ($(root.east -| z.north) + (0, -.5)$) .. (root.east); 
            \draw[densely dotted, -latex, red] (sum1.north) .. controls ($(root.west -| sum1.north) + (0, -.5)$) .. (root.west); 
            
            \draw[densely dotted, -latex, red] (prod1.north) .. controls ($(sum1.west -| prod1.north) + (0, -.5)$) .. (sum1.west);
            \draw[densely dotted, -latex, red] (prod2.north) .. controls ($(sum1.west -| prod2.north) + (0, -.5)$) .. (sum1.east);

            \draw[densely dotted, -latex, red] (x1.north) .. controls ($(prod1.west -| x1.north) + (0, -.5)$) .. (prod1.west);
            \draw[densely dotted, -latex, red] (y1.north) ..controls ($(prod1.east -| y1.north) + (0, -.5)$) .. (prod1.east);

            \draw[densely dotted, -latex, red] (x2.north) .. controls ($(prod1.west -| x2.north) + (0, -.5)$) .. (prod2.west);
            \draw[densely dotted, -latex, red] (y2.north) ..controls ($(prod1.east -| y2.north) + (0, -.5)$) .. (prod2.east);

            \node at ($(root.north) + (0, .2)$) [prob] {$\frac{1}{3} P(Z)$};
            
            \node at ($(root.west -| sum1.north) + (0, -.5)$) [prob, above left] {$\frac{1}{3}$};
            \node at ($(root.east -| z.north) + (0, -.5)$) [prob, above right] {$P(Z)$};
            
            \node at ($(sum1.west -| prod1.north) + (0, -.5)$) [prob, above left] {$\frac{1}{2} \times \frac{2}{3}$};
            \node at ($(sum1.east -| prod2.north) + (0, -.5)$) [prob, above right] {$0 \times \frac{1}{3}$};
            
            \node at ($(prod1.west -| x1.north) + (0, -.5)$) [prob, above left] {$\frac{1}{2}$};
            \node at ($(prod1.east -| y1.north) + (0, -.5)$) [prob, above right] {1};
            
            \node at ($(prod2.west -| x2.north) + (0, -.5)$) [prob, above left] {0};
            \node at ($(prod2.east -| y2.north) + (0, -.5)$) [prob, above right] {1};
        \end{tikzpicture}
        \subcaption{Inferring from the Sum-Product Network by combining probabilities from the leaf nodes. Sum nodes add probabilities, normalized to a valid probability by their weights. Product nodes multiply probabilities, which are assumed to be independent.}
        \label{subfig:infer-spn}
    \end{subfigure}                                                  
    \caption{Sum-Product Network learning and inference.}
    \label{fig:spn}
\end{figure*}
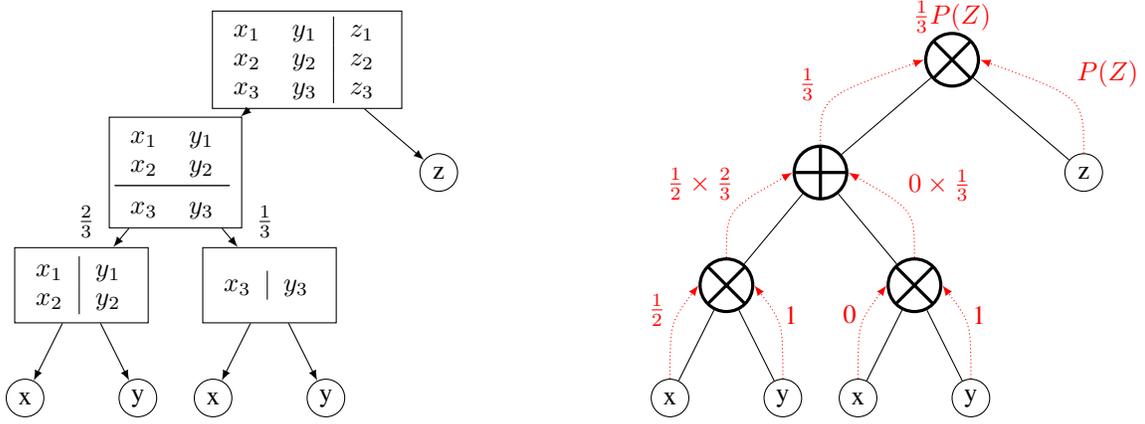

\subsection{Sum-Product Networks}
\label{prelim:spn}
Our model choice, Sum-Product Networks~\cite{spn} (SPNs) are probabilistic graphical models defined as a rooted acyclic graph --- a tree.
Each \textit{leaf} is an independent random variable represented by a probability density function (PDF).
The root and any internal node are either a \textit{sum} or \textit{product} node.
Every sum node is a mixture of PDFs.
Every product node is a product of PDFs.
Each node is considered a valid PDF for the joint probability distribution of its descendants --- a sum node can be a mixture of product nodes, and a product node can be a product of sum nodes.
Thus, the root represents the joint PDF of all random variables in an SPN.

\subsubsection{Structure Learning}
SPNs can decompose complex joint probability distributions as linear combinations of simpler probability distributions.
Gens and Domingos~\cite{learn-spn} offer a simple recursive algorithm to learn a tree structure for SPNs.
Briefly, it initializes the SPN from the root, and grows the tree by recursively checking three cases:
\begin{enumerate}
    \item If the data is a single random variable, i.e., an attribute, return a leaf node and terminate.
    \item Else, if the data can be decomposed into independent groups of random variables, return a product node whose children are those groups.
    \item Otherwise, partition the data by clustering similar tuples and return a sum node.
\end{enumerate}
\noindent
Each new child recursively checks these cases until terminating with a leaf.
This generic algorithm does not assume how independent groups or clusters are determined.

\subsubsection{Example}

\autoref{subfig:learn-spn} depicts an example of learning an SPN for a table with columns $X$, $Y$, and $Z$, whose elements are not necessarily distinct. Starting top-down, only $Z$ is determined to be independent, thus forming a leaf node. The remaining $(X, Y)$ tuples form a sum node containing two clusters, weighted by their proportion of tuples. $X$ and $Y$ are locally independent within both clusters, terminating into leaf nodes. Since every branch has terminated in a leaf node, the SPN structure learning process is complete.

\autoref{subfig:infer-spn} illustrates the inference of joint probabilities from the SPN, e.g., for each value of $Z$ with the predicate $X=x_2$.
For this example, let $x_2$ be distinct from $x_1$ and $x_3$.
Starting bottom-up, each leaf of $X$ returns $P(X=x_2)$.
$Y$ has no predicates and is marginalized out by its leaves returning a probability of 1.
These probabilities of $X$ and $Y$ are multiplied at product nodes.
At the sum node, the normalized sum of probabilities is $P(X=x_2)=\frac{1}{3}$.
The root multiplies the PDF $P(Z)$ with $P(X=x_2)$, expressing the subset of $P(X, Z)$ where $X=x_2$. 
Substituting $P(Z)$ for a frequency distribution (sketch) approximates the frequency distribution (sketch) of the selection where $X=x_2$.

\subsubsection{Application to Relational Data}
Molina and Vergari et al.~\cite{mspn} proposed learning SPNs with the Randomized Dependence Coefficient (RDC) metric~\cite{rdc}, which non-linearly transforms random variables.
This produces a type-agnostic feature space suitable for testing independence and clustering mixed data types, i.e., both continuous and discrete.
The ability to simultaneously handle different data types is highly relevant for relational data, which may contain multiple attribute types.

Hilprecht et al.~\cite{deepdb} applies SPNs as approximate query processors for relational databases.
Additionally, they show that SPNs may be efficiently updated --- inserting a tuple simply requires finding and updating a relevant subset of nodes.
For join cardinality estimation, their best accuracy is obtained by modeling the full outer join of multiple relations.
However, our results show that SPNs can achieve high accuracy without the full outer join, which is intractable for large databases.

\section{Sketched Sum-Product Networks}

Inspired by the success of Fast-AGMS~\cite{conv-sketch} and Bound Sketch~\cite{pessimistic}, we use SPNs to approximate these sketches.
This allows sketches to be used without the need to scan relations and apply filter conditions during estimation time.
We start by training an ensemble of SPNs, one per relation.
After training, each SPN can approximate a sketch for any selection, given its filter conditions.
These sketches are then used for the subsequent join cardinality estimation task.

\subsection{Training Sketched SPNs}

Our algorithm for modeling multivariate data, i.e., a relation, and training an SPN is given in \autoref{alg:train-SPN}, which generally follows the recursive template by Gens and Domingos~\cite{learn-spn} that was described in \autoref{prelim:spn}.
We modify the termination case: if the current partition of data is a single attribute or only contains join attribute(s) then we create a leaf that stores the sketch of the attribute(s).
A leaf may contain multiple join attributes, such that it stores a multivariate sketch, i.e., a sketch tensor or a convolved sketch for cross-correlation.
The other recursive cases remain: if possible, decompose attributes into independent groups, or else partition the data into clusters.

\begin{algorithm}[htb]
\caption{TrainSPN}\label{alg:train-SPN}
\begin{algorithmic}
    \Require relation $T$ with attribute set $\{C\}$
    \Ensure Sum-Product Network of relation $T$

    \If{$\left|C\right| = 1$}
        \State \Return univariate sketch of $\{C\}$
    \ElsIf{$C$ contains only join attributes}
        \State \Return multivariate sketch of $\{C\}$
    \Else
        \State $\{G\} \gets$ decompose $\{C\}$ into independent groups
        \If {$\left|G\right| > 1$}
            \State \Return $\prod_i \mathbf{TrainSPN}(\{G_i\})$
        \Else
            \State $P \gets$ partition $T$ into clusters of similar tuples
            \State \Return $\sum_i \frac{\left|P_i\right|}{\left|R\right|} \mathbf{TrainSPN}(P_i)$
        \EndIf
    \EndIf
\end{algorithmic}
\end{algorithm}

As proposed by Molina and Vergari et al.~\cite{mspn}, product nodes utilize the RDC metric~\cite{rdc} to measure pairwise independence.
Briefly, the RDC metric randomly transforms each attribute and evaluates its linear correlation within a non-linear type-agnostic feature space.
This is applicable between different attribute types, i.e., continuous and discrete.
Dependent attributes form connected components, where RDC less than a user-specified threshold indicates independence.
Each component becomes the child of a product node and recursively calls \autoref{alg:train-SPN}.
If attributes cannot form separable components, then a sum node is created instead.

Sum nodes partition the tuples into clusters, with the goal of forming clusters that have locally independent attributes.
A sum node partitions data into exactly two clusters, as recommended to create deeper networks\footnote{Delalleau and Bengio~\cite{deep_spn} proved deeper SPNs are more efficient than shallow SPNs at modeling certain functions. Experiments by Vergari et al.~\cite{simplifying-spns} corroborate this.}.
Following prior work~\cite{mspn,deepdb}, we originally applied K-Means clustering to the same non-linear features utilized by the RDC test.
However, we found that EM~\cite{em}, as originally recommended for SPNs~\cite{spn, learn-spn}, is more effective for maximizing the independence between attributes within the same cluster.
This ultimately leads to faster training, since fewer partitions are needed before forming leaf nodes.

A leaf node is created whenever an attribute is pairwise independent of all other attributes within its partition.
Alternatively, if the number of tuples within a partition is less than some percentage (e.g., $1\%$) of the original relation, then a leaf node is created for each attribute.
These user-specified thresholds control the termination cases of \autoref{alg:train-SPN} and limit the size of the model.

\subsubsection{Sketches in the Leaf Nodes}
Leaf nodes represent the distribution of attribute(s) using sketches.
These sketches combine to approximate the sketch of any given selection.
This is also viable for any \textit{mergeable synopses}~\cite{Cormode:tods-2013:mergeable-summaries} in general, but hash-based sketches (e.g., Fast-AGMS and Bound Sketch) are particularly suitable since their bins are inherently aligned by hash functions.
This allows them to combine under simple element-wise operations.

\subsection{Inferring Sketches}

\autoref{alg:inference} is applied to the root of a Sketched SPN and recursively traverses the network depth-first to infer the sketch of the given join attribute(s) $\mathcal{K}$ for the selection $\sigma_{\varphi}(T)$.
The predicate $\varphi$ may contain disjunctive (i.e., \textit{condition OR condition}) and conjunctive conditions (i.e., \textit{condition AND condition}).

\begin{algorithm}[htb]
\caption{ApproxSketch}\label{alg:inference}
\begin{algorithmic}
    \Require SPN node $V$ with children $\left\{V_i\right\}$,\\
        selection predicate $\varphi$,\\
        join attributes(s) $\mathcal{K}$
    \Ensure Sketch of attribute(s) $\mathcal{K}$ from the selection $\sigma_\varphi$
    \If{$V$ is a leaf node}
        \If{$\mathcal{K} \subseteq$ the attributes of $V$}
            \State \Return sketch of $\mathcal{K}$
        \Else
            \State \Return selectivity $P\left(\varphi\right) \in [0,1]$
        \EndIf
    \ElsIf{$V$ is a product node}
        \State \Return $\prod_i \mathbf{ApproxSketch}\left(V_i, \varphi, \mathcal{K}\right)$
    \ElsIf{$V$ is a sum node}
        \If{$\mathcal{K} \subseteq$ the attributes of $V$}
            \State \Return $\sum_i \mathbf{ApproxSketch}\left(V_i, \varphi, \mathcal{K}\right)$
        \Else
            \State \Return $\sum_i \frac{\left|V_i\right|}{\left|V\right|} \mathbf{ApproxSketch}\left(V_i, \varphi, \mathcal{K}\right)$
        \EndIf
    \EndIf
\end{algorithmic}
\end{algorithm}

Upon reaching a leaf, the recursive function returns a sketch of join attribute(s) $\mathcal{K}$, if $\mathcal{K}$ is represented by the leaf.
Otherwise, the leaf returns its estimated selectivity for the predicate $\varphi$, which is the probability $P(\varphi) \in [0, 1]$.
If the leaf attribute(s) are excluded from the conditions in $\varphi$, then the returned selectivity is 1, i.e., the attribute(s) are marginalized out.

Sketches and probabilities from the leaf nodes are combined bottom-up by sum and product nodes until a sketch is returned at the root.
This assumes sketches are \textit{linear}~\cite{synopses}, such that adding the sketches of two relations equals the sketch of their union.
However, this does not hold for the degree component of the Bound Sketch method, which is non-linear.
The sum of the maximum degrees from different multisets may overestimate the maximum degree of their union.
Hence, the Bound Sketch upper bound may be looser when approximated by SPNs.
This can be alleviated by using the maximum degree of the whole multiset to constrain the approximation:
\begin{equation}
    \widehat{\mathcal{D}}\left(\sigma_\varphi\left(T\right)\right) \leq
    \mathcal{D}\left(T\right)
\end{equation}
\noindent
where the left side of the inequality is the approximated maximum degree sketch for the selection $\sigma_\varphi\left(T\right)$.
The right side is the exact maximum degree sketch of the unfiltered relation $T$, assumed to be available at estimation time.

We originally attempted to modify sum nodes to merge degree sketches by taking their element-wise maximum.
Ideally, if the elements partitioned into different leaf nodes were distinct, then the largest maximum degree of each leaf node equals the maximum degree of their union.
However, this assumption often did not hold and the resulting estimates tended to underestimate.
Rather than enforce this assumption, it is simpler to treat it as linear and allow the summation and multiplication of degree sketches.
Doing so is still tighter than the Count-Min upper bound:
\begin{equation}
    \lvert A \Join B \rvert \leq
    \mathcal{C}(A) \cdot \left(\mathcal{D}\left(B_1\right) + \mathcal{D}\left(B_2\right)\right)
    \leq
    \mathcal{C}(A) \cdot \mathcal{C}(B)                    
\end{equation}
\noindent
where $B_1$ and $B_2$ are disjoint subsets of the relation $B$.
For all possible partitions of $B = B_1 \bigcup B_2$, the sum of the maximum degree sketches $\mathcal{D}(B_1)$ and $\mathcal{D}(B_2)$ is never greater than the Count-Min 
sketch $\mathcal{C}(B)$.
They are only equal in the worst case that all elements inserted into a bin share the same value. 

\subsubsection{Support for Predicates}
The estimator at the leaf node must handle the various conditions in the selection predicate $\varphi$.
Currently, the sketches in this work apply to equalities, as well as ranges treated as disjunctive equalities.
Estimation of range selectivity is optimized by sketching the \textit{dyadic intervals}~\cite{dyadic} containing an element, as proposed by Cormode and Muthukrishnan~\cite{count-min}.
Dyadic intervals are intervals whose sizes are powers of 2, e.g., $2^0, 2^1, 2^2,$ etc.
Any range of size $n$ can be decomposed into $\mathcal{O}(\log_2n)$ disjoint dyadic intervals, which are treated as disjunctive equalities for estimation.
Future work may include different synopses to support additional predicates.
Only synopses for leaf nodes of join attributes gain from being mergeable element-wise, e.g., sketches.

\subsection{Error Bounds}

Two assumptions affect approximation error: (1) leaf nodes can accurately estimate selectivity, and (2) the children of a product node are independent.
The first is resolved by improving the individual estimators in the leaf nodes, e.g., by allocating more memory.
The second requires minimizing the dependence between attributes, e.g., tightening the RDC or cluster size threshold.
However, these increase the model's memory requirements and training time.
Therefore, it is useful to know when an SPN is ineffective, before retraining.
We can determine its effectiveness by checking whether its error is near its worst-case.

\begin{conjecture}
    The absolute error of any approximated sketch counter is at most the error of fully assuming independence between attributes and scaling the sketch of the join attribute(s) by the exact selectivity of predicate $\varphi$ on each attribute $r \in T$.
    \label{th:error-bounds}
\end{conjecture}

We formally express the error bound as the L1-distance between the exact and approximate sketch upper bounded by the L1-distance between the exact and worst-case approximate sketch. 
The worst-case approximation is by a single product node or a complete independence assumption between attributes.
\begin{equation}
    \biggl\| \mathcal{S}\left(\sigma_\varphi(T)\right) 
            - \widehat{\mathcal{S}}(\sigma_\varphi(T))
    \biggr\|_1  \leq 
    \biggl\| \mathcal{S}\left(\sigma_\varphi(T)\right) 
            - \mathcal{S}(T) \prod_{r \in T} P(\varphi_r)
    \biggr\|_1
    \label{eq:error-bound}
\end{equation}
\noindent
where the function $\mathcal{S}(T): \mathds{R}^{N \times M} \rightarrow \mathds{R}^w$ is a linear mapping from a relation $T$ to a $w$-dimensional sketch, e.g., Fast-AGMS.  
Then $\widehat{\mathcal{S}}(T)$ denotes an approximate sketch.
This worst-case error bound assumes that the exact selectivity of predicates is given by each leaf node.
As such, the actual worst-case error bound may be higher in practice, and error should be measured over many queries.

\subsection{Join Cardinality Estimation}

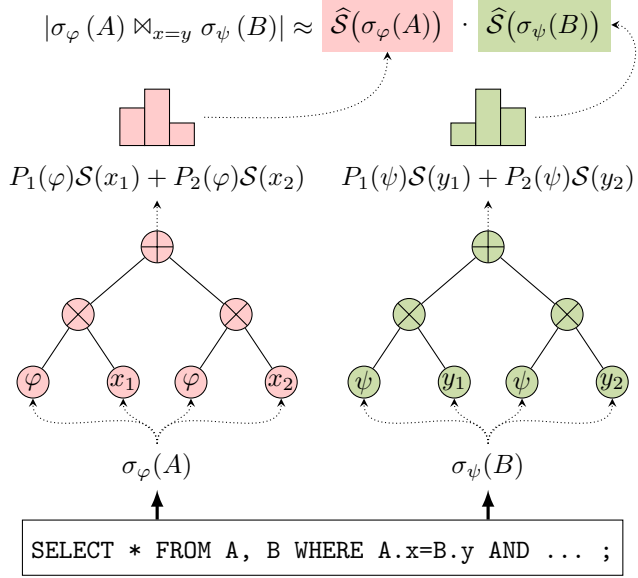
\begin{figure}[th]
    \centering
    \begin{tikzpicture}
        [level distance = 0.75cm,
        level 1/.style = {sibling distance=1.75cm},
        level 2/.style = {sibling distance=1cm},
        sum/.style = {draw, circle, align=center,
            path picture={\draw
                (path picture bounding box.south) -- (path picture bounding box.north)
                (path picture bounding box.west) -- (path picture bounding box.east);}},
        product/.style = {draw, circle, align=center, 
            path picture={\draw
                (path picture bounding box.south east) -- (path picture bounding box.north west)
                (path picture bounding box.south west) -- (path picture bounding box.north east);}
        },
        edge from parent/.style = {draw, -},
        scale=1.2]

        \node[draw] (sql) {
            \begin{lstlisting}
SELECT * FROM A, B WHERE A.x=B.y AND ... ;
            \end{lstlisting}
        };

        \node[anchor=south] at ($(sql.north west)!0.45!(sql.north) + (0, 0.33)$) (sigma_A) {
            $\sigma_\varphi(A)$
        };

        \node[anchor=south] at ($(sql.north east)!0.45!(sql.north) + (0, 0.33)$) (sigma_B) {
            $\sigma_\psi(B)$
        };

        \draw[-{latex}, very thick] 
            ($(sigma_A.south |- sql.north)$)
            to [out=90, in=-90] (sigma_A.south);
        \draw[-{latex}, very thick] 
            ($(sigma_B.south |- sql.north)$) 
            to [out=90, in=-90] (sigma_B.south);


        \node[sum, fill=palered, inner sep=0.15cm, anchor=south] 
            at ($(sigma_A.north) + (0, 2)$) (spn_A){}
            child {node[product, fill=palered, inner sep=0.15cm] {}
                child {node[draw, circle, fill=palered, inner sep=0.15cm, label=center:$\varphi$] (leaf_a11){}}
                child {node[draw, circle, fill=palered, inner sep=0.15cm, label=center:$x_1$] (leaf_a12){}}
            }
            child {node[product, fill=palered, inner sep=0.15cm] {}
                child {node[draw, circle, fill=palered, inner sep=0.15cm, label=center:$\varphi$] (leaf_a21){}}
                child {node[draw, circle, fill=palered, inner sep=0.15cm, label=center:$x_2$] (leaf_a22){}}
            }
        ;

        \node[sum, fill=palegreen, inner sep=0.15cm, anchor=south]
            at ($(sigma_B.north) + (0, 2)$) (spn_B){}
            child {node[product, fill=palegreen, inner sep=0.15cm] {}
                child {node[draw, circle, fill=palegreen, inner sep=0.15cm, label=center:$\psi$] (leaf_b11){}}
                child {node[draw, circle, fill=palegreen, inner sep=0.15cm, label=center:$y_1$] (leaf_b12){}}
            }
            child {node[product, fill=palegreen, inner sep=0.15cm] {}
                child {node[draw, circle, fill=palegreen, inner sep=0.15cm, label=center:$\psi$] (leaf_b21){}}
                child {node[draw, circle, fill=palegreen, inner sep=0.15cm, label=center:$y_2$] (leaf_b22){}}
            }
        ;

        \draw[densely dotted, -{stealth}, thin] (sigma_A) to [out=90, in=-90] (leaf_a11);
        \draw[densely dotted, -{stealth}, thin] (sigma_A) to [out=90, in=-90] (leaf_a12);
        \draw[densely dotted, -{stealth}, thin] (sigma_A) to [out=90, in=-90] (leaf_a21);
        \draw[densely dotted, -{stealth}, thin] (sigma_A) to [out=90, in=-90] (leaf_a22);
        \draw[densely dotted, -{stealth}, thin] (sigma_B) to [out=90, in=-90] (leaf_b11);
        \draw[densely dotted, -{stealth}, thin] (sigma_B) to [out=90, in=-90] (leaf_b12);
        \draw[densely dotted, -{stealth}, thin] (sigma_B) to [out=90, in=-90] (leaf_b21);
        \draw[densely dotted, -{stealth}, thin] (sigma_B) to [out=90, in=-90] (leaf_b22);

        \node[anchor=south] at ($(spn_A.north) + (0, 0.33)$) (expr_A) {
            $P_1(\varphi)\mathcal{S}(x_1) + P_2(\varphi)\mathcal{S}(x_2)$
        };
        \node[anchor=south] at ($(spn_B.north) + (0, 0.33)$) (expr_B) {
            $P_1(\psi)\mathcal{S}(y_1) + P_2(\psi)\mathcal{S}(y_2)$
        };

        \draw[densely dotted, -{stealth}] (spn_A.north) to [out=90, in=-90] (expr_A.south);
        \draw[densely dotted, -{stealth}] (spn_B.north) to [out=90, in=-90] (expr_B.south);

        \node[anchor=south] at ($(expr_A.north) + (0, 0)$) (sketch_A) {
            \begin{tikzpicture}
                \draw[fill=palered] (0,0) rectangle (0.33,0.5);
                \draw[fill=palered] (0.33,0) rectangle (0.66,0.75);
                \draw[fill=palered] (0.66,0) rectangle (0.99,0.3);
            \end{tikzpicture}
        };
        \node [anchor=south] at ($(expr_B.north) + (0, 0)$) (sketch_B) {
            \begin{tikzpicture}
                \draw[fill=palegreen] (0,0) rectangle (0.33,0.3);
                \draw[fill=palegreen] (0.33,0) rectangle (0.66,0.75);
                \draw[fill=palegreen] (0.66,0) rectangle (0.99,0.5);
            \end{tikzpicture}
        };

        \node[matrix, anchor=south] at ($(sketch_A.north)!0.5!(sketch_B.north) + (0, 0.2)$) (join_card) {
            \node{$\lvert \sigma_\varphi\left(A\right)
            \Join_{x=y}
            \sigma_\psi\left(B\right)\rvert
            \approx$}; &
            \node[fill=palered] (shat_A) {$\widehat{\mathcal{S}}\big(\sigma_\varphi(A)\big)$}; &
            \node[right=of shat_A]{$\cdot$}; &
            \node[fill=palegreen] (shat_B)
            {$\widehat{\mathcal{S}}\big(\sigma_\psi(B)\big)$}; \\
        };

        \draw[densely dotted, -{stealth}] (sketch_A.east) to [out=0, in=-90] (shat_A.south); 
        \draw[densely dotted, -{stealth}] (sketch_B.east) to[out=0, in=-90] ($(expr_B.east |- shat_B.south east) + (-0.1, 0)$)
        to[out=90, in=0] (shat_B.east);

    \end{tikzpicture}
    \caption{Join cardinality estimation using SPNs to approximate the sketches of the join keys $x$ and $y$ from their selections $\sigma_\varphi(A)$ and $\sigma_\psi(B)$, respectively. The dot product of these sketches is a join cardinality estimate.}
    \label{fig:estimation}
\end{figure}

The process of inferring sketches for join cardinality estimation is exemplified in \autoref{fig:estimation}.
It depicts two SPNs, one for a relation $A$ and another for relation $B$.
They share the same structure: the root is a sum node that partitions the data into two clusters forming product nodes. 
These product nodes terminate into leaf nodes for a join attribute and a generic selection predicate attribute. 
Consider the two-way join $\sigma_\varphi(A) \Join_{x=y} \sigma_\psi(B)$, where $x$ and $y$ are the join attributes of $A$ and $B$ respectively.
To estimate the size of the join, each SPN is used to infer the sketch of the join keys within their respective selection. 

Given the selection predicate $\varphi$, the SPN of $A$ combines each cluster's probability of satisfying $\varphi$ with its local sketch of the join attribute $x$.
Let $P_1(\varphi)$ and $\mathcal{S}(x_1)$ denote the probability and sketch for the first cluster, while $P_2(\varphi)$ and $\mathcal{S}(x_2)$ belong to the second cluster.
Then, the SPN expresses the sketch of the selection as the sum of products:
\begin{equation}
    \widehat{\mathcal{S}}\left(\sigma_\varphi\left(A\right)\right)
    =
    P_1\left(\varphi\right) \mathcal{S}\left(x_1\right)
    + P_2\left(\varphi\right) \mathcal{S}\left(x_2\right)
\end{equation}
\noindent
The approximate sketch of $\sigma_\psi(B)$ is similarly expressed in terms of probabilities and sketches.

After inferring the sketch of each selection, join cardinality estimation follows the original process for each sketch method.
In general, the dot product of the inferred sketches estimates the cardinality of the two-way join, requiring that the sketching function $\mathcal{S}$ of the inferred sketches share the same hash function(s).
Typically, multiple estimates are taken for better accuracy, using different hash functions to create multiple independent sketch estimators.
The median trick is applied for unbiased Fast-AGMS estimates.
For the pessimistic Bound Sketch, the minimum of its estimates is returned as a tight upper bound.


\subsubsection{Probabilistic Upper Bound}
\label{subsec:upward-bias}

Estimators that guarantee an upper bound on the actual join cardinality are referred to as \textit{pessimistic}~\cite{pessimistic, pessimistic-cardest}.
However, even without a strong guarantee, upward biased estimators have still been shown~\cite{factorjoin} to benefit query optimization, sometimes even more so than exact cardinality from an oracle~\cite{elephant}.
Intuitively, overestimating cardinality encourages query optimizers to plan more cautiously and tends towards plans that are only marginally suboptimal, whereas allowing for underestimation increases the risk of choosing plans with potentially catastrophic execution times.

To induce an upward bias in our estimator, we take the maximum of multiple Fast-AGMS estimates\footnote{We also used the more stable upper quartile estimate, but found that the maximum estimate produced marginally better query execution times.}, instead of the median.
Furthermore, we modify the product node to use the minimum univariate selectivity, rather than the product, to scale the sketch of the join attribute.
Since the minimum univariate selectivity is an upper bound on the product, this allows more of any sketch it multiplies with to \textit{survive} the approximation.
Henceforth, we refer to this modification as the \textit{min-product node}.
The min-product node further increases the upward bias of the maximum Fast-AGMS estimate, and is also applicable to Bound Sketch.
In our experiments, we evaluate the efficacy of these upward biasing techniques for query optimization.

\section{Experiments}

\subsection{Implementation}

We implement Sketched SPNs in Python, as are our comparisons.
Sketches are stored in a sparse tensor format~\cite{sparse}, which may prevent the model size from increasing linearly with the size of the sketches in their leaf nodes.
Sketches are materialized at estimation time, since sparse operations may be significantly slower than the same dense operations.
For selectivity estimation in leaf nodes, we use the simple Count-Min sketch, which only requires a single hash function for simpler and faster inference. 
We use the $k$-universal hash function~\cite{universal-hash} implemented by Heddes et al.~\cite{conv-sketch} to construct sketches.

\subsection{Setup}

Experiments are executed on an Ubuntu 24 system with an Intel Xeon E5-2660 v4 CPU and 256 GB RAM.
Specifically, for query execution time, we use the modified PostgreSQL 13.1 provided by Han and Wu et al~\cite{stats-ceb}, which implements commands to plug in cardinality estimates from external methods.
It also disables parallel workers for query execution, which emphasizes the impact of the cardinality estimator on query execution speed.

\subsection{Datasets}

We evaluate on the JOB-light~\cite{mscn} and Stats-CEB~\cite{stats-ceb} workloads, which are commonly used to evaluate join cardinality estimation methods.
Han and Wu et al~\cite{stats-ceb} report that the attributes in Stats-CEB are more skewed and correlated with each other.
Hence, its data is expected to be more complex and difficult to model accurately.

\begin{itemize}
    \item JOB-light consists of 70 join queries (696 subqueries) on 6 relations from IMDb~\cite{IMDB}.
    These queries are transitive joins on up to 5 relations in a star schema.
    \item Stats-CEB consists of 146 join queries (2603 subqueries) on 8 relations from Stats Stack Exchange data~\cite{stackexchange}.
    It includes non-transitive joins on up to 7 relations.
\end{itemize}

A subquery is a subset of joins from the original query, along with relevant selection predicates.
In order to execute a query, the cardinality estimate of each subquery is passed to PostgreSQL.
We also evaluate the accuracy of join cardinality estimation methods, using these subqueries.

\begin{figure}[th]
    \centering
    \includegraphics[width=0.8\linewidth]{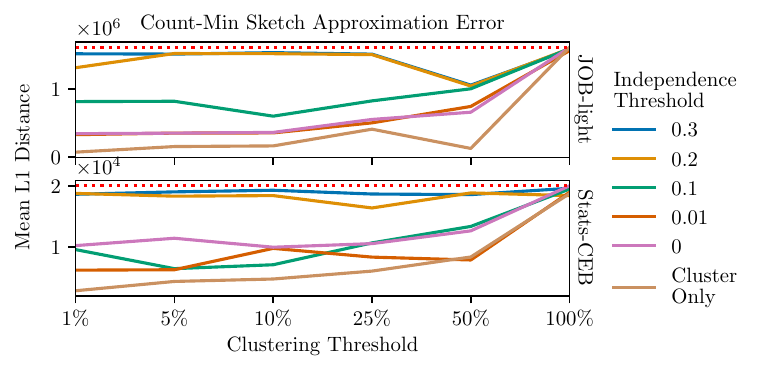}
    \caption{Mean L1-distance between the exact Count-Min sketch of a selection and its approximation by SPNs with varied complexity.
    The worst-case independence assumption model gives the upper bound (dashed line) on error.
    }
    \label{fig:l1-bound}
\end{figure}

\subsection{Sketch Approximation Error}
We verify our error bound (\autoref{eq:error-bound}) and evaluate how closely SPNs can approximate sketches.
Since it is only applicable to linear sketches, we evaluate the Count-Min sketch components of the Bound Sketch method.
Notably, the sum of the counters in the Count-Min sketch equals the cardinality of its selection.
Hence, the Count-Min sketch approximation error is analogous to the SPN's cardinality estimation error for a single-table query.

We compute the L1-distance between an exact Count-Min sketch and its SPN approximation, for each selection in our workloads.
When a query does not specify filter conditions for a relation --- a selection --- then its sketch returned by the SPN is simply the sum of sketches from its leaf nodes.
This is equivalent to the exact sketch of the unfiltered relation.
We omit such sketches from our evaluation, since their error is 0. 
There are 1\,165 and 5\,451 selections specified in the JOB-light and Stats-CEB workloads, respectively.

\autoref{fig:l1-bound} evaluates SPNs of various complexities, by adjusting the minimum clustering size and independence thresholds.
As these thresholds decrease, the SPN is expected to become increasingly accurate, which improves the sketch approximation error.

The minimum clustering size is specified as a percentage of the original relation.
A clustering threshold of 100\% means the data is never partitioned by a sum node and is just a single product node --- the worst-case complete independence assumption model, which gives the upper bound on our sketch approximation error.
We tighten the clustering threshold as low as 1\%, meaning the SPN only forms sum nodes on partitions that no smaller than $1\%$ of the original relation.
However, the SPN still might not completely partition the relation until reaching this threshold, unless the attributes within all partitions are insufficiently independent.

\begin{figure}[ht]
    \centering
    \includegraphics[width=0.8\linewidth]{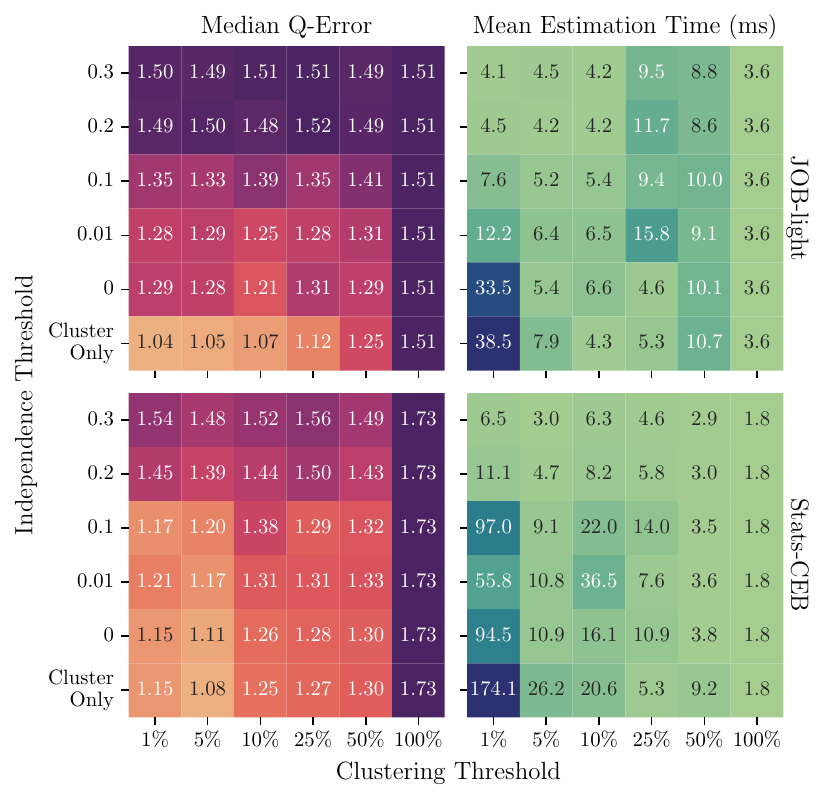}
    \caption{Q-error and estimation time for JOB-light and Stats-CEB workloads, using Fast-AGMS sketch approximations.
    By tightening training thresholds, q-error decreases in a trade-off with estimation time.}
    \label{fig:heatmaps}
\end{figure}

The RDC metric measures non-linear dependency between two attributes and has the range $[0, 1]$.
An independence threshold of $0$ means that a product node is only formed whenever an attribute is pairwise independent of all other attributes in the same instance.
We find that an independence threshold of at least $0.2$ fails to decrease approximation error.
This is regardless of the clustering threshold, since the independence threshold is so large that the minimum clustering size is never met.
We suggest that it is more practical to set the independence threshold to 0 and primarily control model complexity using the clustering threshold.

Eschewing the independence condition, e.g., setting the threshold below 0, means that the SPN repeatedly forms clusters via sum nodes until each cluster reaches the clustering threshold.
Product nodes are only made afterwards.
This is a special case of the model --- equivalent to simply clustering --- and the approximation becomes the sum of sketches made by assuming independence locally within each cluster.
\autoref{fig:l1-bound} shows that this is more accurate than SPNs of the same clustering threshold with a non-negative independence threshold.
It even appears to serve as a lower bound on the sketch approximation error. 
Although the simplicity of only clustering is attractive, the objective of using SPNs is to achieve similar accuracy with a smaller model.

\subsection{Join Cardinality Estimation Accuracy}

In join cardinality estimation, accuracy is often measured using \textit{q-error}~\cite{qerror}, defined as the largest ratio between a positive cardinality $Y$ and its estimate $\widehat{Y}$:
\begin{equation}
    \operatorname{q-error} = \max \left\{\frac{\widehat{Y}}{Y}, \frac{Y}{\widehat{Y}}\right\}
\end{equation}
\noindent
\autoref{fig:heatmaps} analyzes the effect of the SPN training thresholds on the q-error of their approximate Fast-AGMS sketches, using sketch width $w=10^5$ and taking the median of 5 independent estimates.
Since Bound Sketch is not unbiased, its q-error is higher than Fast-AGMS' and we omit it.
However, the SPN hyperparameters similarly affect the approximation error of either sketch, which correlates with their q-error on join cardinality.

Generally, q-error decreases as we tighten either threshold.
The independence threshold must be as small as 0, in order to improve over the worst-case independence assumption model corresponding to a 100\% clustering threshold. 
This is especially for JOB-light, where a high independence threshold only marginally improves q-error.
In contrast, decreasing the clustering threshold until 10\% quickly improves q-error.

Estimation time does not increase significantly with each smaller clustering threshold, until it is 1\%.
Instead, an SPN with fewer clusters (via a higher clustering threshold) may sometimes be slower. 
This is an effect of the sparse tensor format~\cite{sparse} used in our implementation --- operations on sketches with more non-zero elements are slower than sparser sketches.
It is not guaranteed that the sketches corresponding to smaller clusters would be sparser.
Until it is 0, the independence threshold also has little apparent impact on estimation time. 

\begin{figure}[th]
    \centering
    \includegraphics[width=0.8\linewidth]{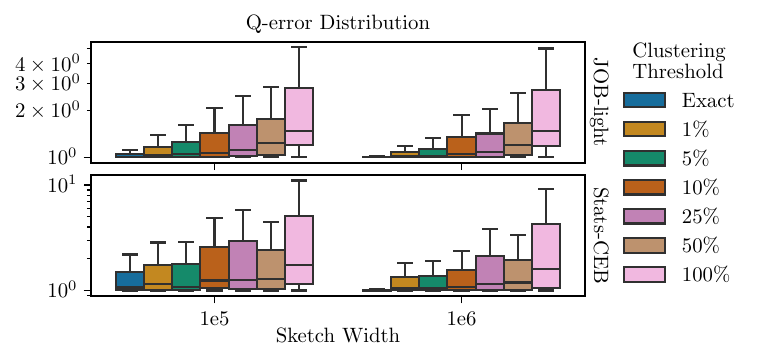}
    \caption{Q-error distribution of exact Fast-AGMS sketches and their approximations by SPNs with various clustering thresholds. The independence threshold is fixed to 0.}
    \label{fig:qerror_by_width}
\end{figure}

We also compare the distribution of q-errors for exact and approximate Fast-AGMS sketches.
Our objective is for the approximate sketches' q-errors to approach that of exact sketches.
In \autoref{fig:qerror_by_width}, we verify that tightening the clustering threshold of SPNs used to approximate Fast-AGMS sketches also tightens the distribution of their q-error.
However, it falls short of exact sketches.
The smallest 1\% clustering threshold is excessive, since it causes high estimation time with marginal benefits to q-error.
A larger width also only marginally improves q-error on our datasets.

\subsection{Sketching Efficiency}

Ideally, the computational time and space requirements for the approximation process should be less than exact sketching, to be considered practical.
Computing an exact sketch entails scanning\footnote{Following prior work~\cite{compass, conv-sketch}, we simply implement linear scanning, but it is also possible to utilize index scans for faster exact sketching~\cite{pessimistic}.} and filtering a relation, hashing each element that satisfies the selection, and updating the sketch.
Since hashing and updating are simple operations, the bulk of the work is in scanning and filtering.
We assume an idealized scenario for exact sketches: that each distinct sketch (i.e., for a particular selection) is only computed once and saved for reuse.
This reduces the time and space requirements for exact sketching, making a more nuanced comparison with our approximations.

For all of the sketches required for our workloads, \autoref{tab:sketch_efficiency} reports the time to compute them, either exactly or approximately.
For exact sketching, both Fast-AGMS (shortened to F-AGMS) and Bound Sketch, we report the total space required to store sketches sparsely.
For approximations, denoted as F-AGMS\textsuperscript{\textdagger} and Bound Sketch\textsuperscript{\textdagger}, we report the SPN ensemble size.
SPNs are trained with an independence threshold of 0 and a clustering threshold of 10\%, which balances acceptable accuracy and estimation time for our workloads.
Henceforth, these are the training hyperparameters in our other experiments.  

\begin{table}[th]
    \centering
    \caption{Computational requirements of exact sketches and approximations (denoted with \textdagger) using width $10^5$.}
    \begin{tabular}{l r r r r}
        \toprule
        {} & \multicolumn{2}{c}{JOB-light} & \multicolumn{2}{c}{Stats-CEB} \\ 
        Sketch & Time (s) & Space & Time (s) & Space \\  
        \midrule
        F-AGMS & 737.7 & 472.6 MB & 154.6 & 3.1 GB \\ 
        F-AGMS\textsuperscript{\textdagger} & 4.5 & 1.2 GB & 17.3 & 675.7 MB \\ 
        Bound Sketch & 520.1 & 279.9 MB & 233.7 & 2.5 GB \\ 
        Bound Sketch\textsuperscript{\textdagger} & 35.7 & 671.9 MB & 59.5 & 343.9 MB \\ 
        \bottomrule
    \end{tabular}
    \label{tab:sketch_efficiency}
\end{table}

The time required to approximate both Fast-AGMS and Bound Sketch is always faster than exact sketching.
Unlike exact sketching, approximate sketches are not saved for reuse --- an SPN must approximate each sketch, every time it is needed.
This is a practical scenario that does not assume a priori knowledge of the selections in the query workloads, which may be prohibitive.
Nonetheless, approximation via SPN is at least a few factors faster than exact sketching.
In particular, Fast-AGMS is up to two orders of magnitude faster to approximate, since the exact version requires multiple hash functions.
Bound Sketch requires fewer hash functions, but uses both Count-Min and degree sketches.

The space requirement of approximate sketching is higher than exact sketching on JOB-light, but vice versa on Stats-CEB, which contains more selections.
JOB-light contains few enough unique selections that the total size of sketches saved for those selections is smaller than the SPN.
The opposite is true for Stats-CEB.
This suggests that for workloads with few unique selections, exact sketches may be more practical.

\begin{table}[ht]
    \centering
    \caption{Training time for the ensembles of \textit{data-driven} and \textit{learned} cardinality estimators.}
    \begin{tabular}{l c c c}
         \toprule
         {} & {} & \multicolumn{2}{c}{Training Time (min)} \\
         Method & Model & JOB-light & Stats-CEB \\
         \midrule
         DeepDB~\cite{deepdb} & SPN & 29.91 & 54.95 \\
         Sketched SPNs & SPN & 20.86 & 2.11 \\
         BayesCard~\cite{bayescard} & Bayes net & 5.61 & 1.78 \\
         FactorJoin~\cite{factorjoin} & Bayes net & 3.91 & 0.52 \\
         \bottomrule
    \end{tabular}
    \label{tab:training-time}
\end{table}

\subsection{Model Training Time}

Our method is also comparable to other \textit{learned} cardinality estimators, specifically \textit{data-driven} methods.
Data-driven cardinality estimators observe the tuples of the target relation to model its distribution. 
This is in contrast to \textit{query-driven} cardinality estimators, which train on queries annotated with their ground-truth cardinality~\cite{mscn, fauce, alece}.

The training time of Sketched SPNs and other data-driven learned cardinality estimators is given in \autoref{tab:training-time}.
These are categorized as ensembles of either SPNs or Bayesian networks~\cite{bayes-net}.
Like this work, DeepDB~\cite{deepdb} also applies an ensemble of SPNs to model relations and estimate join cardinality.
However, each SPN in DeepDB may be trained on (a sample of) either one relation or the full outer join of multiple.
BayesCard~\cite{bayescard} uses Bayesian networks instead.
FactorJoin~\cite{factorjoin} also uses Bayesian networks, but eschews full outer joins and constrains each network to a single relation.
We train these models using their default or recommended hyperparameters, if any are given.

DeepDB and BayesCard both limit the number of full outer joins used, which can be costly.
However, their training time is still higher than their per-relation counterparts for the same model.
In particular, the full outer join of the relations in Stats-CEB is four orders of magnitude larger than JOB-light's, as reported by Han and Wu et al.~\cite{stats-ceb}.
Thus, DeepDB has a much longer training time on Stats-CEB than Sketched SPNs.

SPNs require noticeably longer training than Bayesian networks.
One reason is that the K-means clustering method, as used in DeepDB, may fail to effectively cluster data into locally independent attributes during training --- an SPN may repeatedly create sum nodes until reaching the clustering threshold.
However, we observe that using Hard Expectation-Maximization (EM)~\cite{em} for clustering, instead of K-means, prevents this.

\begin{table}[ht]
    \centering
    \caption{Impact of clustering method on training SPNs.}
    \begin{tabular}{l c c}
        \toprule
         Workload & Cluster Method & Structure Learning \\ 
         \midrule
         \multirow{2}{4.5em}{JOB-light} 
            & K-means & 5 min 22 sec \\ 
            & Hard EM & 3 min 29 sec \\ 
         \midrule 
         \multirow{2}{4.5em}{Stats-CEB} 
            & K-means & 1 min 2 sec \\ 
            & Hard EM & 1 min 19 sec \\ 
         \bottomrule
    \end{tabular}
    \label{tab:clustering}
\end{table}

Structure learning time of SPNs using either clustering method is compared in \autoref{tab:clustering}.  
Structure learning refers to forming the SPN nodes via clustering and independence testing.
On JOB-light, K-means fails to form clusters that minimize the dependency between attributes, whereas Hard EM results in 35\% faster structure learning.
On the other hand, K-means has no issues on Stats-CEB and Hard EM offers no improvement.
We observe little difference to accuracy nor model size.
Although K-means may be faster, Hard EM is recommended to be cautious.

\subsection{Query Execution Time}

\begin{figure*}[th]
    \centering
    \includegraphics[width=\linewidth]{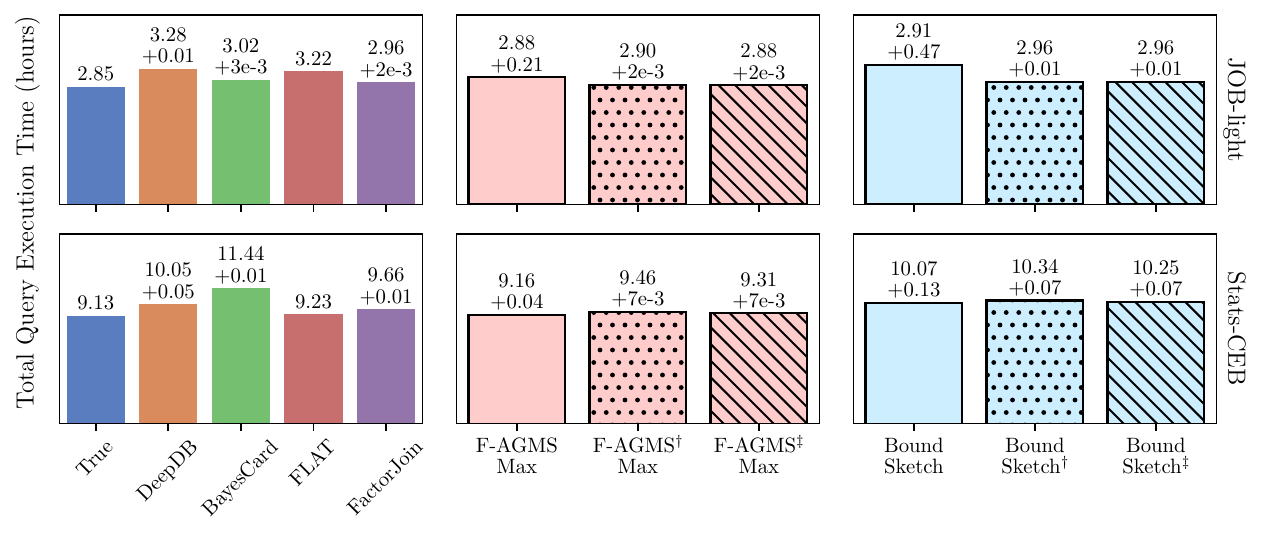}
    \caption{Total end-to-end query execution times, which includes  the added cardinality estimation times.
    SPN sketch approximations are denoted by \textdagger.
    Approximations by SPNs using our min-product node are denoted by \textdaggerdbl~ instead.}
    \label{fig:query_times}
\end{figure*}

\autoref{fig:query_times} shows the total end-to-end query execution times (including cardinality estimation time) of the different estimators and also the ground truth cardinality.
We include the data-driven cardinality estimator by Zhu and Wu et al., FLAT~\cite{flat}, which proposes an SPN variant called the Factorize-Split-Sum-Product Network (FSPN).
The FSPN identifies highly correlated attributes and factorizes their joint probability distribution into conditional probability distributions, e.g., as multivariate histograms.
This efficiently models the attributes that our SPN would otherwise struggle to decompose into independent univariate distributions.
At the time of writing, FLAT does not have an open-source implementation that supports join cardinality estimation.
However, its estimates for JOB-light and Stats-CEB are provided by Han and Wu et al.~\cite{stats-ceb}.

For each sketch method, we evaluate their exact sketches and our approximations all using width $10^5$.
The maximum of Fast-AGMS estimates (F-AGMS Max) is used instead of its unbiased median.
We find that the median estimate is uncompetitive on Stats-CEB, unless a larger sketch width is used, as Heddes et al.~\cite{conv-sketch} showed with a width of $10^6$ --- the median estimate still resulted in slower query execution than FLAT on Stats-CEB.
In comparison, Fast-AGMS Max has query execution time that is second only to the ground-truth cardinality.

Although the approximations do not attain as fast query execution as exact sketches, they compensate with their lower estimation time.
In particular, our min-product node approximations, Fast-AGMS\textsuperscript{\textdaggerdbl} Max and Bound Sketch\textsuperscript{\textdaggerdbl}, demonstrate the effectiveness of upward-biased estimation in query optimization.
Note that the estimation time for exact sketches includes their construction time --- we do not assume that the sketches were already available.
This would require knowing the necessary sketches to prepare beforehand.
Otherwise, the estimation time for exact sketches would be under a minute.

Overall, Fast-AGMS Max results in faster query execution than Bound Sketch.
This is unexpected, since Bound Sketch guarantees overestimation.
On the other hand, Fast-AGMS Max may still underestimate, which is commonly cited~\cite{pessimistic, elephant} as riskier than overestimation to cause sub-optimal query execution plans.

\subsection{Relative Error Distribution}


\begin{figure}[th!]
    \centering
    \includegraphics[width=0.8\linewidth]{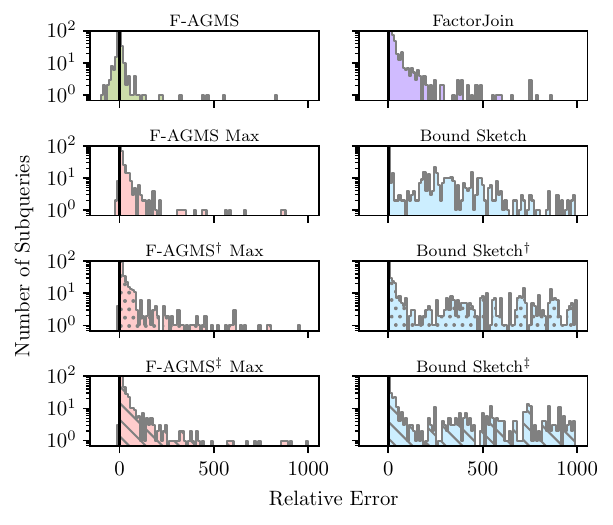}
    \caption{Distribution of relative errors on Stats-CEB}
    \label{fig:rel_error}
\end{figure}

We analyze the bias of our estimators in \autoref{fig:rel_error}, which shows the distribution of the relative estimation errors for Fast-AGMS Max and Bound Sketch.
It also includes the unbiased Fast-AGMS median estimator to verify that Fast-AGMS Max is significantly more upward-biased.
However, the heavy-tailed distribution of relative errors for Bound Sketch reveals that its upward bias is much stronger.
It may greatly overestimate even thousands of times the actual join cardinality.
In comparison, Fast-AGMS Max and FactorJoin --- another upward-biased estimator --- are still highly accurate.
Although it is not guaranteed, they effectively produce tighter upper bounds, thus achieving faster query execution.
We refer to Bergmann et al.~\cite{elephant} for an analysis of the impact of overestimating cardinalities.

\subsection{Approximate Sketches}

This work closely mirrors Approximate Sketches~\cite{ApproximateSketches}, prior work that trained bidirectional transformers~\cite{bert} to also approximate Fast-AGMS sketches.
It shares the same premise --- an ensemble of per-relation models is trained to approximate the sketch of any selection whose filter conditions are given at estimation time.
Unlike SPNs, bidirectional transformers are dependent on hardware accelerators (e.g., GPU) for training.
As such, the size of models that can be trained is limited by the accelerator's memory.
Furthermore, the model size grows linearly with the sketches to approximate --- each counter in a sketch has a trainable embedding.
Thus, Approximate Sketched used a relatively small width of up to 4\,096.
The join cardinality estimator was a heuristic~\cite{compass} that allows sketches with fewer hash functions , thus fewer that may need to be approximated, but is restricted to transitive joins, e.g., JOB-light.
We compare it to Fast-AGMS\textsuperscript{\textdaggerdbl} Max estimator, using the same sketch width of 4\,096, in \autoref{tab:approxsketch}.

\begin{table}[th]
\centering
    \caption{Comparison to prior work, Approximate Sketches, which is only implemented for transitive joins (e.g., JOB-light) and a smaller sketch width of up to 4096.
    *The time of a single epoch is given for Approximate Sketches.}
    \begin{tabular}{l c c}
        \toprule
        Method & \shortstack{Approximate\\ Sketches~\cite{ApproximateSketches}} & Sketched SPNs \\
        \midrule
        Models & BERT~\cite{bert} & SPN~\cite{spn} \\
        Model size & 167 MB & 40 MB \\
        Training time & 148 min* & 18 min \\
        Estimator & Heuristic~\cite{compass} & F-AGMS\textsuperscript{\textdaggerdbl} Max \\
        Mean estimation time & 13 ms & 11 ms \\
        Median q-error & 3.36 & 1.57 \\
        Total query execution &  3.05 + \num{3e-3} hrs & 2.93 + \num{2e-3} hrs \\
        \bottomrule
    \end{tabular}
    \label{tab:approxsketch}
\end{table}

The training time of Approximate Sketches is reported for a single epoch of gradient descent.
However, even a single epoch (on an NVIDIA Tesla K80 GPU) exceeds our SPN training time.
Approximate Sketches was trained to approximate 5 independent sketch estimators, which we found we needed to double to 10 for Fast-AGMS\textsuperscript{\textdaggerdbl} Max.
This was due to the increased variance of a smaller width sketch producing a few extreme underestimations that affected just 4 out of the 70 queries in JOB-light, but resulted in a 0.2 hrs longer total execution time than Approximate Sketches.
Using more independent estimators prevents such underestimations.
Thus Fast-AGMS\textsuperscript{\textdaggerdbl} Max achieves faster query execution befitting its lower q-error.

\section{Related Work}

\subsection{Sketches for Join Cardinality Estimation}
The Fast-AGMS sketch~\cite{fast-agms} was derived from the AGMS sketch proposed for unbiased join cardinality estimation by Alon et al~\cite{agms}.
AGMS can be seen as a special case of Fast-AGMS with a width of one --- a single counter.
In practice, a large number of independent AGMS estimators were necessary for accuracy.
However, its update time increases linearly with the number of counters.
Fast-AGMS~\cite{count-sketch, fast-agms} attains sub-linear update time by partitioning data into an array of multiple counters.

To the best of our knowledge, AGMS sketches were first applied to join cardinality estimation subject to filter conditions by Vengerov et al.~\cite{vengerov}.
At query time, any given selection predicate can be treated as a join with a \textit{virtual} relation containing all values that satisfy the predicate.
The AGMS sketches of these virtual relations could be computed on-the-fly to estimate that join.
This might seem inefficient when virtual relations represent large ranges, but the ad hoc AGMS sketch can be calculated analytically for certain hash functions~\cite{fast_random_variables}.
Thus, any join with filter conditions is treated as one between join relations and virtual relations altogether.
However, treating selection predicates as joins greatly increases estimation variance~\cite{multi-way-agms}.

Ganguly et al.~\cite{dense-collisions} observed that join cardinality estimation error is largely due to collisions with frequent elements.
They proposed \textit{skimming}~\cite{skimmed-sketches} the frequent elements from sketches into separate estimators, which involves iterating over the domain of inserted elements.
Roy et al.~\cite{augmented-sketch} avoid iteration by tracking and preemptively filtering frequent elements.
Wang et al.~\cite{JoinSketch} also separately store the elements that are not yet determined to be either frequent or infrequent.
Join cardinality estimation with these \textit{multifocal} methods requires estimating the join between every combination of partitions. 
In the \textit{bifocal} case, we decompose $\lvert A \Join B \rvert$ into 4 combinations: the frequent elements of $A$ with the frequent elements $B$, the frequent elements of $A$ with the infrequent elements of $B$, the infrequent elements of $A$ with the frequent elements of $B$, and the infrequent elements of $A$ with the infrequent elements of $B$.
By storing the counts of frequent elements more accurately (e.g., exactly) their collisions can be reduced.
However, the number of combinations increases exponentially with the number of joins, which makes it impractical for our workloads.
\balance 

\subsection{Learned Cardinality Estimators}

Yang et al.~\cite{neurocard} proposed NeuroCard, another data-driven learned cardinality estimator.
They trained a deep autoregressive model~\cite{made, attention} on (a sample of) a full outer join.
Autoregressive models predict the (conditional) probability distribution of an attribute subject to specific values of other attributes, i.e., given by filter conditions.
Thus, they model the full outer join's joint probability distribution factorized as a product of conditional probabilities, e.g., $P(X, Y) = P(X \vert Y) P(Y)$.
Join cardinality is estimated as the sum of probabilities --- each for a valid join tuple $(X, Y)$ --- sampled from the full outer join.

Kim et al.~\cite{ASM} also use deep autoregressive models~\cite{made} but in an ensemble of per-relation models, rather than a full outer join.
They perform join cardinality estimation by importance sampling~\cite{importance-sampling} of probabilities from the deep autoregressive models.
In an ablation study with FactorJoin, which uses another sampling method~\cite{adaptive-sampling}, they show that importance sampling is generally more effective, regardless of the per-relation model type.
Future work may compare sampling to sketching for the estimation of join cardinalities from ensembles.
We suggest that composing an ensemble estimate via sampling may enable the use of smaller models, whereas high-dimensional sketches have better accuracy. 

Query-driven cardinality estimators~\cite{mscn, fauce} train on queries labeled with their ground-truth cardinality.
Naturally, the training workload should be representative of the testing workload.
Furthermore, estimators may be invalidated by dynamically shifting distributions of the underlying data.
Hybrid methods address this by using both data-driven and query-driven training~\cite{unified-data-query-ce} or incorporating the database state as an input~\cite{robust-query-driven-ce, alece}.
Relevantly, Liu et al.~\cite{qspn} recently proposed a query-aware Sum-Product Network (QSPN) that incorporates query-driven training into the construction of the SPN.
Exceptionally, they use an unlabeled query workload to determine the \textit{access affinity} between attributes --- the frequency that attributes are referenced in the same query together.
Instead of just pairwise independent attributes, those with low access affinity can also have their joint probability distribution factorized by a product node.
Although we limit our scope to an orthodox SPN, such variants (e.g., FSPN and QSPN) may improve sketch approximation.

\section{Conclusions}

We propose Sketched Sum-Product Networks that can approximate an ad hoc sketch for any selection with arbitrary filter conditions.
This provides an alternative pipeline to the standard assumption that sketches are constructed and maintained for queries known a priori, which limits their applications.
We implement both Fast-AGMS and Bound Sketch methods and show that their approximations via SPNs can sufficiently match both their accuracy and query optimization efficacy.
In particular, we show that the maximum of Fast-AGMS estimates is an effective upward-biased join cardinality estimator that achieves fast query execution time on both JOB-light and Stats-CEB.
Since SPNs can approximate sketches very quickly, they do not incur the cost of constructing sketches from scratch.
On workloads with many distinct selections, it is also more memory-efficient to approximate with SPNs than maintain exact sketches.
While future work may explore models besides SPNs, our results are highly promising for the effectiveness of incorporating sketches into ensembles of cardinality estimators.
Our work is also applicable to other linear sketches that may even be applied to other tasks than cardinality estimation.

\section*{Acknowledgements}
This work was supported by NSF award number 2008815.

\bibliographystyle{abbrv}

\end{document}